\documentclass[11pt,a4paper]{article}
\usepackage[english]{babel}
\usepackage{amsmath,amsthm,amssymb,epsfig,latexsym}
\usepackage{color}
\usepackage{ulem}
\newcommand{\so}{\scriptscriptstyle \rm I}
\newcommand{\st}{\scriptscriptstyle \rm I\hspace{-1pt}I}
\newcommand{\sth}{\scriptscriptstyle \rm I\hspace{-1pt}I\hspace{-1pt}I}

\newcommand{\as}{\lambda}
\newcommand{\bla}{\bar u}
\newcommand{\bmu}{\bar v}
\newcommand{\blac}{\bar{u}^{\scriptscriptstyle C}}
\newcommand{\blab}{\bar{u}^{\scriptscriptstyle B}}
\newcommand{\bmuc}{\bar{v}^{\scriptscriptstyle C}}
\newcommand{\bmub}{\bar{v}^{\scriptscriptstyle B}}
\def\E{{\sf E}}
\def\Uq#1{U_q(\widehat{\mathfrak{gl}}_{#1})}
\def\Uqin#1{U_{q^{-1}}(\widehat{\mathfrak{gl}}_{#1})}

\def\Izer{{\sf K}}
\def\bbb{\mathbb{B}}
\def\ccc{\mathbb{C}}
\newcommand{\Izerl}{\Izer^{(l)}}
\newcommand{\Izerr}{\Izer^{(r)}}
\newcommand{\Izerlr}{\Izer^{(l,r)}}
\newcommand{\Izerrl}{\Izer^{(r,l)}}
\newcommand{\RHCl}{{\sf Z}^{(l)}}
\newcommand{\RHCr}{{\sf Z}^{(r)}}
\newcommand{\RHClr}{{\sf Z}^{(l,r)}}

\newcommand{\be}[1]{\begin{equation}\label{#1}}
\newcommand{\ba}[1]{\begin{multline}\label{#1}}
\newcommand{\ee}{\end{equation}}
\newcommand{\ea}{\end{eqnarray}}

\newcommand{\num}{\\\rule{0pt}{20pt}}

\newcommand{\tr}{\mathop{\rm tr}}

\newtheorem{prop}{Proposition}[section]
\newtheorem{lemma}{Lemma}[section]

\def\qed{\hfill\nobreak\hbox{$\square$}\par\medbreak}
 \makeatletter
 \@addtoreset{equation}{section}
 \makeatother
 
\begin{document}
\markright{\today\dotfill \jobname\dotfill }

\def\AQ{\mathcal{A}_q}
\def\AQin{\mathcal{A}_{q^{-1}}}

\begin{flushright}
LAPTH-004/14
\end{flushright}

\vspace{20pt}

\begin{center}
\begin{LARGE}
 {\bf Scalar products in  models with $GL(3)$ \\[1.2ex]
 trigonometric $R$-matrix. General case\\[1.2ex]
 }
\end{LARGE}

\vspace{40pt}

\begin{large}
{S.~Pakuliak${}^a$, E.~Ragoucy${}^b$, N.~A.~Slavnov${}^c$\footnote{pakuliak@theor.jinr.ru, eric.ragoucy@lapth.cnrs.fr, nslavnov@mi.ras.ru}}
\end{large}

 \vspace{12mm}

${}^a$ {\it Laboratory of Theoretical Physics, JINR, 141980 Dubna, Moscow reg., Russia,\\
Moscow Institute of Physics and Technology, 141700, Dolgoprudny, Moscow reg., Russia,\\
Institute of Theoretical and Experimental Physics, 117259 Moscow, Russia}

\vspace{4mm}

${}^b$ {\it Laboratoire de Physique Th\'eorique LAPTH, CNRS and Universit\'e de Savoie,\\
BP 110, 74941 Annecy-le-Vieux Cedex, France}

\vspace{4mm}

${}^c$ {\it Steklov Mathematical Institute,
Moscow, Russia}

\vspace{4mm}


\end{center}

\vspace{2mm}

\begin{abstract}
We study quantum integrable models with
$GL(3)$ trigonometric $R$-matrix solvable by the nested algebraic Bethe ansatz. We analyze scalar
products of generic Bethe vectors and obtain an explicit representation for them in terms of a sum with respect to partitions of Bethe parameters. This representation generalizes known formula for the scalar products in the models with
$GL(3)$-invariant $R$-matrix.
\end{abstract}

\vspace{2mm}


\renewcommand*{\thefootnote}{\arabic{footnote}}
\addtocounter{footnote}{-1}

\section{Introduction}

The algebraic Bethe ansatz \cite{FadST79,FadLH96,KulS79,FadT79} allows one to obtain the spectrum of quantum Hamiltonians in many models of physical interest. Calculation of correlation functions  also can be performed in the framework
of this method. The last problem in many cases can be reduced to the calculation of scalar products of Bethe
vectors. In our previous work \cite{PakRS13c} we began a systematic study of scalar products in  quantum integrable models with
$GL(3)$ trigonometric $R$-matrix. In that paper we have calculated the highest
coefficients of the scalar product and found their properties. Using these results we continue the investigation of
the scalar products in the present paper.

For the $GL(2)$-based models the study of scalar products of Bethe vectors in the framework of the
algebraic Bethe ansatz was initiated in the works \cite{Kor82,IzeKor84,Kor84,BogIK93L}. There it was used
for the calculation of the correlation functions in the models of one-dimensional bosons and $XXZ$ Heisenberg
chain. In the work \cite{Sla89} a determinant representation for the scalar product of an arbitrary Bethe vector
and an eigenvector of the transfer matrix was obtained. This result allowed later to obtain multiple
integral representations for correlation functions in
various quantum integrable models \cite{KitMT00,KitMST02,KitKMST09b,GohKS04,GohKS05,SeeBGK07}. The determinant
representation for the scalar product also was used for the calculation of form factors of local operators
\cite{Sla90,IzeKMT99,KitKMST11a}. These results were used for the analytical \cite{KitKMST11,KitKMST12} and numerical analysis
of correlation functions \cite{CauHM05,PerSCHMWA06,PerSCHMWA07,CauCS07}.

Formally, the scalar product of Bethe vectors should play the same role
in the models described by the higher rank algebras. However this case is much more sophisticated from
the technical viewpoint. Up to now the only representation for the scalar product for the models with
$GL(3)$-invariant $R$-matrix is known due to N. Reshetikhin \cite{Res86}. This representation was given in terms of
a sum with respect to partitions of Bethe parameters (so-called sum formula). Such type of formulas are not appropriate neither for
analytical nor for numerical calculations. However, they can be used to produce more compact formulas, in particular, to find determinant representation for some particular cases of  the scalar products. Representations of this type were obtained
in \cite{Whe12,BelPRS12b,BelPRS13a} for some important particular cases of the scalar products in quantum integrable  models with
$GL(3)$-invariant $R$-matrix.

The question arises about the generalization of the Reshetikhin representation to the trigonometric case.
In distinction of the $GL(2)$-based models this generalization is not straightforward. In particular,
in the  paper \cite{PakRS13c} we have shown that in the models with  $GL(3)$ trigonometric $R$-matrix
there exist two highest coefficients, while in the case of $GL(3)$-invariant $R$-matrix there is only one highest coefficient.

The main goal of this paper is to derive such the generalization. In the paper \cite{PakRS13c} we have made the
first step on this way. Now we go further and obtain an analog of the Reshetikhin formula for the scalar product for the models with  $GL(3)$ trigonometric $R$-matrix. We use the same method as in \cite{PakRS13c}. It is based on the explicit representations for the dual Bethe vectors and formulas of multiple action of the monodromy matrix entries onto Bethe vectors. This method is straightforward, although it is a bit technical.  In many respects it relies on the properties of the highest coefficients established in \cite{PakRS13c}.

The content of the paper is as follows. In section~\ref{sec:bckgr} we describe the model under consideration and introduce
necessary notations. In  section~\ref{S-MR} we present the main result of the paper: the sum formula for the scalar
product of Bethe vectors in the models with $GL(3)$ trigonometric $R$-matrix. The remaining part the paper is devoted to the proof of this formula. In section~\ref{S-NT} we describe the tools that are used for the derivation of the sum formula. In section~\ref{S-SA}
we evaluate a multiple successive action of the monodromy matrix entries onto Bethe vectors. Using this result we
obtain the sum formulas for the scalar product in section~\ref{S-ScalP}. Appendices~\ref{A-PID}~and~\ref{A-PZab}
contain the properties of the Izergin determinant and the highest coefficients. In appendix~\ref{A-PP} we prove
a summation identity for the highest coefficients.

\section{General background\label{sec:bckgr}}%

\subsection{The model}

The $GL(3)$ trigonometric quantum $R$-matrix has the following form
\begin{equation}\label{UqglN-R}
\begin{split}
R(u,v)\ =\ f(u,v)&\ \sum_{1\leq i\leq 3}\E_{ii}\otimes \E_{ii}\ +\
\sum_{1\leq i<j\leq 3}(\E_{ii}\otimes \E_{jj}+\E_{jj}\otimes \E_{ii})
\\
+\ &\sum_{1\leq i<j\leq 3}
(u\,g(u,v) \E_{ij}\otimes \E_{ji}+ v\,g(u,v)\E_{ji}\otimes \E_{ij})\,.
\end{split}
\end{equation}
Here the rational functions $f(u,v)$ and $g(u,v)$   are
\begin{equation}\label{fgg}
f(u,v)=\frac{qu-q^{-1}v}{u-v},\quad g(u,v)=\frac{q-q^{-1}}{u-v}\,,
\end{equation}
where $q$ is a complex number (a deformation parameter), and $(\E_{ij})_{lk}=\delta_{il}\delta_{jk}$, $i,j,l,k=1,2,3$ are $3\times3$ matrices with unit in the intersection of $i$th row and $j$th column and zero matrix elements elsewhere.

In this paper we will consider  quantum integrable models with the $3\times 3$ monodromy matrix
$T(u)$ which satisfies standard commutation relation ($RTT$-relation)
\begin{equation}\label{RTT}
R(u,v)\cdot (T(u)\otimes \mathbf{1})\cdot (\mathbf{1}\otimes T(v))=
(\mathbf{1}\otimes T(v))\cdot (T(u)\otimes \mathbf{1})\cdot R(u,v)\,.
\end{equation}
The entries $T_{ij}(u)$ of the monodromy matrix form the quadratic algebra with commutation
relations given by \eqref{RTT} and act in a quantum space $V$.
 We denote this algebra as $\AQ$ and further will consider certain morphisms of $\AQ$ (see \eqref{phi})
onto $\AQin$.\footnote{If monodromy matrix $T(u)$ is a generating series with respect
to the parameter $u^{-1}$, the algebra $\AQ$ can be identified with the positive Borel subalgebra
in the quantum affine algebra  $\Uq{3}$.}
 We assume that the vector space $V$
 possesses a highest weight vector $|0\rangle\in V$ such that
\begin{equation}\label{rsba}
T_{ij}(u)|0\rangle =0,\quad i>j,\quad T_{ii}(u)|0\rangle= \lambda_i(u)|0\rangle\,,\qquad \lambda_i(u)\in\mathbb{C}[[u,u^{-1}]]\,.
\end{equation}
We also assume that the operators $T_{ij}(u)$ act in a dual space $V^*$ with
a vector $\langle 0|\in V^*$ such that
\begin{equation}\label{drsba}
\langle 0|T_{ij}(u) =0,\quad i<j,\quad  \langle 0|T_{ii}(u)= \lambda_i(u)\langle 0|\,,
\end{equation}
and  $\lambda_i$ are the same as in \eqref{rsba}.

Below we permanently deal with sets of variables and their partitions into subsets. We denote sets of variables by bar: $\bla$, $\bmu$ and so on,
\begin{equation}\label{set111}
\bla = \left\{u_1,\ldots,u_a\right\},\quad \bmu=\left\{v_1,\ldots,v_b\right\}\,.
\end{equation}
If necessary, the cardinalities of the sets will be described in special comments
after the formulas.

If a set of variables is
multiplied by a number $\alpha\bla$ (in particular, $\bla q^{\pm 2}$), then it means that all the
elements of the set are multiplied by this number
\begin{equation}\label{set-numb}
\alpha\bla = \left\{\alpha u_1,\ldots,\alpha u_a\right\},\qquad
\bmu q^{\pm 2}=\left\{v_1q^{\pm 2},\ldots,v_bq^{\pm 2}\right\}\,.
\end{equation}
Union of sets is denoted by braces, for example, $\{\bar w,\bla\}=\bar\eta$. Partitions of sets into disjoint subsets are denoted by
the symbol $\Rightarrow$, and the subsets are numerated by roman numbers. For example, notation $\bar u \Rightarrow \{\bar u_{\so}, \bar u_{\st}\}$ means that the set $\bla$ is divided into two subsets $\bar u_{\so}$ and $\bar u_{\st}$, such that
$\bar u_{\so}\cap \bar u_{\st}=\emptyset$ and $\{\bar u_{\so}, \bar u_{\st}\}=\bla$. Similarly, notation $\bar\eta\Rightarrow
\{\bar\eta_{\rm i},\bar\eta_{\rm ii},\bar\eta_{\rm iii}\}$ means that the set $\bar\eta$ is divided into three subsets
with pair-wise empty intersections and $\{\bar\eta_{\rm i},\bar\eta_{\rm ii},\bar\eta_{\rm iii}\}=\bar\eta$. Sometimes,
when the number of subsets is big, we will use the standard arabic numbers for their numeration.  In such cases we will
give special comments.

Just like in the paper \cite{PakRS13c} we use shorthand notations for  products with respect to sets of
variables:
 \begin{equation}\label{SH-prod}
 T_{ij}(\bar w)=\prod_{w_k\in\bar w}   T_{ij}(w_k);\quad
  \lambda_2(\bar u)=\prod_{u_j\in\bar u}  \lambda_2(u_j);\quad
 r_k(\bar v)= \prod_{v_j\in\bar v} r_k(v_j),
 \end{equation}
where $r_k(w)=\lambda_k(w)/\lambda_2(w)$, $k=1,3$. That is, if the operator $T_{ij}$ or functions $\lambda_i$
and $r_k$ depend on a set of variables, this means that one should
take the product of the operators or the scalar functions with respect to the corresponding set. The same convention will be used for the products of functions $f(u,v)$
\begin{equation}\label{SH-prod1}
 f(u, \bar w)= \prod_{w_j\in\bar w} f(u, w_j);\quad
 f(\bar u,\bar v)=\prod_{u_j\in\bar u}\prod_{v_k\in\bar v} f(u_j,v_k).
 \end{equation}

The central object of the theory of scalar products in the $GL(2)$-based models is the Izergin determinant $\Izer_k(\bar x|\bar y)$
\cite{Ize87}. It also plays an important role in the case of the models with $GL(3)$ trigonometric $R$-matrix. It is defined
for two sets $\bar x$ and $\bar y$ of same cardinality $\#\bar x=\#\bar y=k$:
\begin{equation}\label{Izer}
\Izer_k(\bar x|\bar y)=\frac{\prod_{1\leq i,j\leq k}(qx_i-q^{-1}y_j)}
{\prod_{1\leq i<j\leq k}(x_i-x_j)(y_j-y_i)}
\cdot\det \left[\frac{q-q^{-1}}{(x_i-y_j)(qx_i-q^{-1}y_j)}\right]\,.
\end{equation}
Below we also use two modifications of the Izergin determinant
\begin{equation}\label{Mod-Izer}
\Izerl_k(\bar x|\bar y)= \prod_{i=1}^kx_i\cdot\Izer_k(\bar x|\bar y)\,, \qquad
\Izerr_k(\bar x|\bar y)= \prod_{i=1}^ky_i\cdot\Izer_k(\bar x|\bar y)\,,
\end{equation}
which we call left and right Izergin determinants respectively.
Some properties of the Izergin determinant and its modifications are gathered in appendix~\ref{A-PID}.

Finally, we recall the formulas for the left and right highest coefficients of the scalar products \cite{PakRS13c}.
These rational functions depend on four sets of variables and are denoted as $\RHCl_{a,b}(\bar t;\bar x|\bar s;\bar y)$ and
$\RHCr_{a,b}(\bar t;\bar x|\bar s;\bar y)$ respectively. The subscripts show that
$\#\bar t=\#\bar x=a$ and $\#\bar s=\#\bar y=b$. Both highest coefficients have representations in terms of the
Izergin determinants
 \begin{equation}\label{RHC-IHC-12}
  \RHClr_{a,b}(\bar t;\bar x|\bar s;\bar y)=(-q)^{\mp b}\sum
 \Izerrl_b(\bar s|\bar w_{\so}q^2)\Izerlr_a(\bar w_{\st}|\bar t)
  \Izerlr_b(\bar y|\bar w_{\so})f(\bar w_{\so},\bar w_{\st})\,.
 \end{equation}
 Here $\bar w=\{\bar s,\bar x\}$. The sum is taken with respect to partitions of the set $\bar w\Rightarrow
\{\bar w_{\so},\bar w_{\st}\}$ with $\#\bar w_{\so}=b$ and $\#\bar w_{\st}=a$. Recall also that according to the
convention on the shorthand notations \eqref{SH-prod1} the function $f(\bar w_{\so},\bar w_{\st})$ actually means
the double product of functions $f(\bar w_{j},\bar w_{k})$ over $w_{j}\in \bar w_{\so}$ and $w_{k}\in \bar w_{\st}$.
The superscript $(l,r)$ on  ${\sf Z}_{a,b}$ means that the equation \eqref{RHC-IHC-12} is valid for $\RHCl_{a,b}$ and for $\RHCr_{a,b}$ separately. Choosing  the first or the second component of $(l,r)$ and the corresponding (up or down resp.) exponent of
$(-q)^{\mp b}$ in this equation, we obtain representations either for $\RHCl_{a,b}(\bar t;\bar x|\bar s;\bar y)$  or for $\RHCr_{a,b}(\bar t;\bar x|\bar s;\bar y)$. Some properties of the highest coefficients are given in appendix~\ref{A-PZab}.

\section{Scalar product of Bethe vectors\label{S-MR}}

We recall that the entries of the monodromy matrix act on the space $V$, such that $|0\rangle\in V$. Other
vectors of the space $V$ can be obtained via the successive action of the creation operators
$T_{12}(u)$, $T_{23}(u)$, $T_{13}(u)$  on the vector $|0\rangle$. Among all vectors of this space the
Bethe vectors play the most important role. The procedure to construct the Bethe vectors was
formulated in \cite{KulRes83} in the framework of the nested algebraic Bethe  ansatz (see also
\cite{KulRes81,KulRes82,VT,KP-GLN,EKhP,OPS,PakRS13b}). These vectors depend on complex variables, which are called the Bethe parameters. If the Bethe parameters satisfy the system of Bethe equations (see \cite{KulRes83}), then the
corresponding Bethe vector becomes an eigenstate of the transfer matrix $t(u)=\tr T(u)$.

We denote the Bethe vectors by ${\bbb}^{a,b}(\bla;\bmu)$. In quantum integrable models with a $GL(3)$ trigonometric $R$-matrix
they depend on two sets
of  Bethe parameters $\bla$ and $\bmu$, such that $\#\bla=a$, $\#\bmu=b$ and
$a,b=0,1,\dots$.

Dual Bethe vectors ${\ccc}^{a,b}(\bla;\bmu)$ belong to the dual space $V^*$. They also depend on two sets
of  Bethe parameters $\bla$ and $\bmu$, with $\#\bla=a$, $\#\bmu=b$ and
$a,b=0,1,\dots$.

The scalar products are defined as
\be{Def-SP}
S_{a,b}(\blac;\bmuc|\blab;\bmub)=\ccc^{a,b}(\blac;\bmuc)\bbb^{a,b}(\blab;\bmub)\,,
\ee
where all the Bethe parameters are generic complex numbers. We have added the superscripts $C$ and $B$
to the sets $\bar u$, $\bar v$ in order to stress that the vectors
$\ccc^{a,b}$ and $\bbb^{a,b}$ may depend on different sets of parameters. The main result of this paper
is a sum formula for the scalar product of Bethe vectors.

\begin{prop}\label{Prop-SP}
The scalar product of two Bethe vectors \eqref{Def-SP} is given by
\be{scal-res}
S_{a,b}(\blac;\bmuc|\blab;\bmub)= \sum \frac{r_1(\blac_{\st})r_1(\blab_{\so})r_3(\bmuc_{\st})r_3(\bmub_{\so})}{f(\bmuc,\blac)f(\bmub,\blab)}\,
W_{\text{part}}\begin{pmatrix}\blac_{\st},\blab_{\st},&\blac_{{\so}},\blab_{{\so}}\\
\bmuc_{\so},\bmub_{\so},&\bmuc_{\st},\bmub_{\st}\end{pmatrix}.
\ee
Here the sum runs over all the partitions $\blac\Rightarrow\{\blac_{\so},\blac_{\st}\}$,
$\blab\Rightarrow\{\blab_{\so},\blab_{\st}\}$,  $\bmuc\Rightarrow\{\bmuc_{\so},\bmuc_{\st}\}$ and $\bmub\Rightarrow\{\bmub_{\so},\bmub_{\st}\}$
with $\#\blac_{\so}=\#\blab_{\so}$ and $\#\bmuc_{\so}=\#\bmub_{\so}$.
For a fixed partition with $\#\blac_{\so}=\#\blab_{\so}=k$ and $\#\bmuc_{\so}=\#\bmub_{\so}=n$, (where $k=0,\dots,a$ and
$n=0,\dots,b$), the rational coefficient $W_{\text{part}}$ has the form
 \begin{multline}\label{W-Reshet}
W_{\text{part}}\begin{pmatrix}\blac_{\st},\blab_{\st},&\blac_{\so},\blab_{\so}\\
\bmuc_{\so},\bmub_{\so},&\bmuc_{\st},\bmub_{\st}\end{pmatrix}=f(\blab_{\st},\blab_{\so}) f(\blac_{\so},\blac_{\st})f(\bmub_{\so},\bmub_{\st})f(\bmuc_{\st},\bmuc_{\so})
f(\bmuc_{\so},\blac_{\so}) f(\bmub_{\st},\blab_{\st})\num
\times
\RHCl_{a-k,n}(\blac_{\st};\blab_{\st}|\bmuc_{\so};\bmub_{\so}) \;\RHCr_{k,b-n}(\blab_{\so};\blac_{\so}|\bmub_{\st};\bmuc_{\st})\;.
 \end{multline}
where the highest coefficients $\RHClr$ are given by \eqref{RHC-IHC-12}.
\end{prop}

In the work \cite{PakRS13c} we have found the rational functions $W_{\text{part}}$ corresponding to the
extreme partitions
\be{def:Zlr}
\begin{aligned}
W_{\text{part}}\begin{pmatrix}\blac,\blab,&\emptyset,\emptyset\\
\bmuc,\bmub,&\emptyset,\emptyset\end{pmatrix}&=\RHCl_{a,b}(\blac;\blab|\bmuc;\bmub)\,,\\
W_{\text{part}}\begin{pmatrix}\emptyset,\emptyset,&\blac,\blab,\\
\emptyset,\emptyset,&\bmuc,\bmub\end{pmatrix}&=\RHCr_{a,b}(\blab;\blac|\bmub;\bmuc)\,.
\end{aligned}
\ee
Proposition~\ref{Prop-SP} determines the  functions $W_{\text{part}}$ for arbitrary partitions of the
Bethe parameters. In the following,  we prove proposition~\ref{Prop-SP}.

\section{Necessary tools\label{S-NT}}

In this section we describe the tools that we use for the calculation of the scalar products.
As we have already explained, our method of calculation  is based on the
explicit formulas for the dual Bethe vectors and for the multiple actions of the monodromy matrix entries onto
Bethe vectors. In this way one can obtain a representation for the scalar product as a sum over partitions of
Bethe parameters. In order to simplify this representation we use a special isomorphism between
$\AQ$ and
$\AQin$ algebras.

\subsection{Explicit representations for dual Bethe vectors \label{S-ERDBV}}

We use explicit representations
for dual Bethe vectors ${\ccc}^{a,b}(\bla;\bmu)$, which  were found in \cite{PakRS13b}\footnote{%
In order to reduce \eqref{Or-form}, \eqref{Or-form2} to the representations of \cite{PakRS13b} one should
apply \eqref{K-invers} to the Izergin determinants in these formulas.}
\be{Or-form}
\mathbb{C}^{a,b}(\bla;\bmu) =\sum \frac{(-q)^{k}\Izerl_{k}(\bla_{\so}|q^2\bmu_{\so})}
{\lambda_2(\bmu_{\st})\lambda_2(\bla)f(\bmu_{\st},\bla)}
f(\bmu_{\st},\bmu_{\so})f(\bla_{\st},\bla_{\so})\,
\langle0|T_{32}(\bmu_{\st})T_{31}(\bla_{\so})T_{21}(\bla_{\st}),
\ee
and
\be{Or-form2}
\mathbb{C}^{a,b}(\bla;\bmu) =\sum \frac{(-q)^{-k}\Izerr_{k}(\bla_{\so}|q^2\bmu_{\so})}
{\lambda_2(\bmu)\lambda_2(\bla_{\st})f(\bmu,\bla_{\st})}
f(\bmu_{\so},\bmu_{\st})f(\bla_{\so},\bla_{\st})\,
\langle0|T_{21}(\bla_{\st})T_{31}(\bmu_{\so})T_{32}(\bmu_{\st}).
\ee
Here the sum goes over all partitions of the sets $\bar u\Rightarrow\{\bar u_{\so},\bar u_{\st}\}$ and
$\bar v\Rightarrow\{\bar v_{\so},\bar v_{\st}\}$
such that $\#\bar u_{\so}=\#\bar v_{\so}=k$, $k=0,\dots,\min(a,b)$. Both of  these representations are needed for our purpose.
In section~\ref{S-Morphism} we will show that
\eqref{Or-form} and \eqref{Or-form2} also are  related by an isomorphism
$\varphi$ between $\AQ$ and $\AQin$ algebras.

\subsection{Multiple actions of $T_{ij}$ on Bethe vectors \label{S-MATBV}}

In order to compute the scalar product we need
formulas of the multiple action of the operators $T_{ij}$ with $i>j$ onto the Bethe vectors. These multiple actions
were derived in the work \cite{PakRS13a} (see also \cite{PakRS13c}). We give here the list of necessary formulas,
including some important particular cases.
Below everywhere in this section $\#\bar w =n$, $\{\bar w,\bla\}=\bar\eta$, $\{\bar w,\bmu\}=\bar\xi$.

\paragraph{$\bullet$ Multiple action of $T_{21}$.}
 \begin{multline}\label{A-act21}
 T_{21}(\bar w)\mathbb{B}^{a,b}(\bla;\bmu)=(-q)^n\lambda_2(\bar w)\,\sum
r_1(\bar\eta_{\so})\;
f(\bar\eta_{\st},\bar\eta_{\so})f(\bar\eta_{\st},\bar\eta_{\sth})f(\bar\eta_{\sth},\bar\eta_{\so})
 \frac{f(\bar\xi_{\st},\bar\xi_{\so})}{f(\bar\xi_{\st},\bar\eta_{\so})} \\
 \times  \Izerr_n(q^{-2}\bar w|\bar\eta_{\st})
  \Izerl_n(\bar\eta_{\so}|q^2\bar\xi_{\so})\Izerl_n(\bar\xi_{\so}|q^2\bar w)
 \,\mathbb{B}^{a-n,b}(\bar\eta_{\sth};\bar\xi_{\st}).
 \end{multline}
The sum is taken over partitions of: $\bar\eta\Rightarrow\{\bar\eta_{\so},\bar\eta_{\st},\bar\eta_{\sth}\}$ with $\#\bar\eta_{\so}=\#\bar\eta_{\st}=n$; and
$\bar\xi\Rightarrow\{\bar\xi_{\so},\bar\xi_{\st}\}$ with $\#\bar\xi_{\so}=n$.

\paragraph{$\bullet$ Multiple action of $T_{31}$.}
 \begin{multline}\label{A-act31}
 \hspace{-5mm}T_{31}(\bar w)\mathbb{B}^{a,b}(\bla;\bmu)=\lambda_2(\bar w)\,\sum
r_1(\bar\eta_{\st})\,r_3(\bar\xi_{\so})\;
 \Izerr_n(q^{-2}\bar\eta_{\so}|\bar\xi_{\so})\Izerl_n(\bar\eta_{\st}|q^{2}\bar\xi_{\st})\Izerr_n(q^{-2}\bar w|\bar\eta_{\so})
 \Izerl_n(\bar\xi_{\st}|q^{2}\bar w) \\
 \times
 \frac{f(\bar\eta_{\so},\bar\eta_{\st})f(\bar\eta_{\so},\bar\eta_{\sth})f(\bar\eta_{\sth},\bar\eta_{\st})
 f(\bar\xi_{\so},\bar\xi_{\st})f(\bar\xi_{\so},\bar\xi_{\sth})f(\bar\xi_{\sth},\bar\xi_{\st})}
 {f(\bar\xi_{\so},\bar\eta_{\st})f(\bar\xi_{\so},\bar\eta_{\sth})f(\bar\xi_{\sth},\bar\eta_{\st})}\,
 \mathbb{B}^{a-n,b-n}(\bar\eta_{\sth};\bar\xi_{\sth}).
 \end{multline}
The sum is taken over partitions of:
 $\bar\xi\Rightarrow\{\bar\xi_{\so},\bar\xi_{\st},\bar\xi_{\sth}\}$ with $\#\bar\xi_{\so}=\#\bar\xi_{\st}=n$;
$\bar\eta\Rightarrow\{\bar\eta_{\so},\bar\eta_{\st},\bar\eta_{\sth}\}$ with $\#\bar\eta_{\so}=\#\bar\eta_{\st}=n$.

Remark that in the particular case $a=n$, we have $\bar\eta_{\sth}=\emptyset$, and then
 \begin{multline}\label{A-act31-1}
 \hspace{-5mm}T_{31}(\bar w)\mathbb{B}^{a,b}(\bla;\bmu)=\lambda_2(\bar w)\,\sum
r_1(\bar\eta_{\st})\,r_3(\bar\xi_{\so})\;
 \Izerr_n(q^{-2}\bar\eta_{\so}|\bar\xi_{\so})\Izerl_n(\bar\eta_{\st}|q^{2}\bar\xi_{\st})\Izerr_n(q^{-2}\bar w|\bar\eta_{\so})
 \Izerl_n(\bar\xi_{\st}|q^{2}\bar w)  \\
 \times
 \frac{f(\bar\eta_{\so},\bar\eta_{\st})
 f(\bar\xi_{\so},\bar\xi_{\st})f(\bar\xi_{\so},\bar\xi_{\sth})f(\bar\xi_{\sth},\bar\xi_{\st})}
 {f(\bar\xi_{\so},\bar\eta_{\st})f(\bar\xi_{\sth},\bar\eta_{\st})}\,
 \mathbb{B}^{0,b-n}(\emptyset;\bar\xi_{\sth}).
 \end{multline}

\paragraph{$\bullet$ Multiple action of $T_{32}$.}
 \begin{multline}\label{A-act32}
 T_{32}(\bar w)\mathbb{B}^{a,b}(\bla;\bmu)=
 (-q)^{-n}\lambda_2(\bar w)\,\sum r_3(\bar\xi_{\so})\;
f(\bar\xi_{\so},\bar\xi_{\st})f(\bar\xi_{\so},\bar\xi_{\sth})f(\bar\xi_{\sth},\bar\xi_{\st})
 \frac{f(\bar\eta_{\so},\bar\eta_{\st})}{f(\bar\xi_{\so},\bar\eta_{\st})} \\
+ \times  \Izerr_n(q^{-2}\bar w|\bar\eta_{\so})
 \Izerr_n(q^{-2}\bar\eta_{\so}|\bar\xi_{\so})\Izerl_n(\bar\xi_{\st}|q^2\bar w)
 \,\mathbb{B}^{a,b-n}(\bar\eta_{\st};\bar\xi_{\sth}).
 \end{multline}
The sum is taken over partitions of: $\bar\xi\Rightarrow\{\bar\xi_{\so},\bar\xi_{\st},\bar\xi_{\sth}\}$ with $\#\bar\xi_{\so}=\#\bar\xi_{\st}=n$; and   $\bar\eta\Rightarrow\{\bar\eta_{\so},\bar\eta_{\st}\}$
with $\#\bar\eta_{\so}=n$.

Note that in the special case $a=0$, we have $\bar\eta_{\so}=\bar w$ and $\bar\eta_{\st}=\emptyset$.
If in addition $b=n$, then
$\bar\xi_{\sth}=\emptyset$ and we obtain
 \begin{equation}\label{A-act32-1}
 T_{32}(\bar w)\mathbb{B}^{0,b}(\emptyset;\bmu)=
 \lambda_2(\bar w)\sum r_3(\bar\xi_{\so})
f(\bar\xi_{\so},\bar\xi_{\st})
 \Izerr_n(q^{-2}\bar w|\bar\xi_{\so})\Izerl_n(\bar\xi_{\st}|q^2\bar w)
 \,|0\rangle.
 \end{equation}

{\sl Remark.\label{Rem-card}} In all formulas for the multiple actions, we described the cardinalities of subsets in special comments. Actually these comments are not necessary, since the cardinalities of the subsets are shown explicitly directly in the formulas. Indeed, the subscript of the Izergin determinant indicates the number of elements in both sets of variables on which it depends. Hence, looking, for example, at equation \eqref{A-act21} we conclude that $\#\bar\eta_{\so}=\#\bar\eta_{\st}=\#\bar\xi_{\so}
=\#\bar w=n$. In addition, the superscripts of Bethe vector $\mathbb{B}^{a-n,b}(\bar\eta_{\sth};\bar\xi_{\st})$ show
that $\#\bar\eta_{\sth}=a-n$ and $\#\bar\xi_{\st}=b$. For the reader convenience, below we will continue to give separate comments about the cardinalities of subsets. However  in equations containing a big number of subsets we will skip such descriptions.

\subsection{Isomorphism between $\AQ$ and $\AQin$ algebras\label{S-Morphism}}

In the work \cite{PakRS13b} we have described an isomorphism between positive
Borel subalgebras in the algebras $\Uq{N}$ and $\Uqin{N}$.
We denote this map by $\varphi$. In the case of the algebras $\AQ$ and $\AQin$
 the map $\varphi$ has the form
\be{phi}
\varphi\bigl(T_{i,j}(u)\bigr)=\tilde T_{4-j,4-i}(u),
\ee
where $T_{i,j}(u)\in\AQ$ and $\tilde T_{4-j,4-i}(u)\in\AQin$, respectively. The action of
$\varphi$ on Bethe vectors is given by (see \cite{PakRS13b}):
\be{phi-BV}
\varphi\bigl(\mathbb{B}^{a,b}(\bla;\bmu)\bigr)=\mathbb{B}^{b,a}_{q^{-1}}(\bmu;\bla),\qquad
\varphi\bigl(\mathbb{C}^{a,b}(\bla;\bmu)\bigr)=\mathbb{C}^{b,a}_{q^{-1}}(\bmu;\bla).
\ee
Here we have equipped the Bethe vector and its dual by the additional subscript $q^{-1}$ in order to
stress that these vectors are constructed for the algebra $\AQin$. Generically, the action of
$\varphi$ can be described as follows
\be{phi-F}
\varphi\bigl(\mathcal{F}(T_{i,j}(u);\lambda_i(u);q)\bigr)=
\mathcal{F}(\tilde T_{4-j,4-i}(u);\tilde \lambda_{4-i}(u);q^{-1}\bigr),\quad\text{where}\quad
\tilde\lambda_{i}(u)=\langle0|\tilde T_{ii}(u)|0\rangle.
\ee
Here $\mathcal{F}$ is some polynomial in the operators $T_{ij}(u)$, whose coefficients may also depend on the vacuum eigenvalues
$\lambda_i(u)$ and the parameter $q$.

The map \eqref{phi} is a very powerful tool for the study of the scalar products. We will use it in
section~\ref{S-RedHC1}. Here we use this isomorphism in order to show the equivalence of  the representations
\eqref{Or-form} and \eqref{Or-form2} for
the dual Bethe vectors.

 We start with the representation \eqref{Or-form}, written in $\AQin$:
\be{1phi-1}
\mathbb{C}_{q^{-1}}^{b,a}(\bmu;\bla) =\sum \frac{(-q)^{-k}\Izerl_{k,q^{-1}}(\bmu_{\so}|q^{-2}\bla_{\so})}
{\tilde \lambda_2(\bla_{\st})\tilde \lambda_2(\bmu)f_{q^{-1}}(\bla_{\st},\bmu)}
f_{q^{-1}}(\bla_{\st},\bla_{\so})f_{q^{-1}}(\bmu_{\st},\bmu_{\so})\,
\langle0|\tilde T_{32}(\bla_{\st})\tilde T_{31}(\bmu_{\so})\tilde T_{21}(\bmu_{\st}),
\ee
Here, similarly to \eqref{phi-BV} we have written an additional subscripts $q^{-1}$ to the function
$f$ and the Izergin determinant. This means that $f_{q^{-1}}$ is given by \eqref{fgg} with
$q$ replaced by $q^{-1}$. Similarly $\Izerlr_{k,q^{-1}}$ are given by \eqref{Mod-Izer} and \eqref{Izer},
where $q$ is replaced by $q^{-1}$.

It is easy to check that
\be{q-qm}
f_{q^{-1}}(x,y)=f(y,x), \qquad \Izerlr_{k,q^{-1}}(\bar x|\bar y)=\Izerrl_{k}(\bar y|\bar x).
\ee
Then one can recast \eqref{1phi-1} as follows
\be{1phi-2}
\varphi\bigl(\mathbb{C}^{a,b}(\bla;\bmu)\bigr) =\sum \frac{(-q)^{-k}\Izerr_{k}(q^{-2}\bmu_{\so}|\bla_{\so})}
{\tilde \lambda_2(\bla_{\st})\tilde \lambda_2(\bmu)f(\bmu,\bla_{\st})}
f(\bla_{\so},\bla_{\st})f(\bmu_{\so},\bmu_{\st})\,
\langle0|\tilde T_{32}(\bla_{\st})\tilde T_{31}(\bmu_{\so})\tilde T_{21}(\bmu_{\st}),
\ee
where we have used \eqref{phi-BV}.
Now we apply the mapping $\varphi$ directly to the representation \eqref{1phi-2} using \eqref{phi-F}. Recall that $\varphi$ acts
only on the operators $\tilde T_{ij}$ and the vacuum eigenvalue $\tilde\lambda_2$. Taking into account that $\varphi^2=1$ we obtain
\be{2phi-1}
\mathbb{C}^{a,b}(\bla;\bmu) =\sum \frac{(-q)^{-k}\Izerr_{k}(\bmu_{\so}|\bla_{\so}q^{2})}
{\lambda_2(\bla_{\st})\lambda_2(\bmu)f(\bmu,\bla_{\st})}
f(\bla_{\so},\bla_{\st})f(\bmu_{\so},\bmu_{\st})\,
\langle0|T_{21}(\bla_{\st}) T_{31}(\bmu_{\so})T_{32}(\bmu_{\st}),
\ee
where we used $\Izerr_{k}(q^{-2}\bar x|\bar y)=\Izerr_{k}(\bar x|\bar yq^{2})$ (see \eqref{K-scal}). Thus, we have
reproduced the representa\-tion \eqref{Or-form2}.

\section{Successive action\label{S-SA}}

The first step in the derivation of the scalar product is to find the result of
the multiple successive action of the operators $T_{32}(\bar z)T_{31}(\bar y)T_{21}(\bar x)$ onto
generic Bethe vector $\mathbb{B}^{a,b}(\bla,\bmu)$, where $\bar z$, $\bar y$, $\bar x$, $\bla$, and $\bmu$ are sets
of arbitrary complex numbers. It follows from the explicit representation of the dual Bethe vector \eqref{Or-form} that for our goal it is enough to consider the case  $\#\bar y+\#\bar x=\#\bla$ and $\#\bar y+\#\bar z=\#\bmu$. Therefore we set\footnote{%
 Recall that the cardinalities of the sets $\bla$ and $\bmu$ coincide with the superscripts of the Bethe vector
 $a$ and $b$ respectively.} $\#\bar y=k$, $\#\bar x=a-k$, and $\#\bar z=b-k$, where $k=0,1,\dots,\min(a,b)$.

For the derivation of the multiple successive action we use the formulas \eqref{A-act21}, \eqref{A-act31-1}, \eqref{A-act32-1}.
It is not difficult to guess that the final result will contain a sum over partitions of the original sets of variables
into a big number of subsets. Therefore, in order to avoid cumbersome roman numbers we use in this section standard arabic
numbers for notations of subsets.

\subsection{Successive action of $T_{31}(\bar y)T_{21}(\bar x)$}

Let
\be{S1}
S_1= T_{31}(\bar y)T_{21}(\bar x)\mathbb{B}^{a,b}(\bar u;\bar v).
\ee
The action of $T_{21}(\bar x)$ is given by \eqref{A-act21}, where one should set $\bar w=\bar x$ and $n=a-k$. We
obtain
 \begin{multline}\label{act21}
S_1=(-q)^{a-k}\lambda_2(\bar x)T_{31}(\bar y)\,\sum
r_1(\bar\eta_{1})\;
f(\bar\eta_{2},\bar\eta_{1})f(\bar\eta_{2},\bar\eta_{3})f(\bar\eta_{3},\bar\eta_{1})
 \frac{f(\bar\xi_{2},\bar\xi_{1})}{f(\bar\xi_{2},\bar\eta_{1})} \\
 \times  \Izerr_{a-k}(q^{-2}\bar x|\bar\eta_{2})
  \Izerl_{a-k}(\bar\eta_{1}|q^2\bar\xi_{1})\Izerl_{a-k}(\bar\xi_{1}|q^2\bar x)
 \,\mathbb{B}^{k,b}(\bar\eta_{3};\bar\xi_{2}),
 \end{multline}
where $\eta=\{\bar u,\bar x\}$ and $\xi=\{\bar v,\bar x\}$. Recall that the sum is taken over partitions $\bar\eta\Rightarrow\{\bar\eta_{1},\bar\eta_{2},\bar\eta_{3}\}$ and
$\bar\xi\Rightarrow\{\bar\xi_{1},\bar\xi_{2}\}$. The
cardinalities of the subsets are equal to the subscripts of the corresponding Izergin determinants
or to the superscripts of the Bethe vector (see Remark on  page~\pageref{Rem-card}).

The action of $T_{31}(\bar y)$ creates new
partitions $\{\bar y,\bar\eta_3\}\Rightarrow\{\bar\eta_4,\bar\eta_5\}$ and $\{\bar y,\bar\xi_2\} \Rightarrow\{\bar\xi_3,\bar\xi_4,\bar\xi_5\}$. It means, in particular, that the products over subset
$\bar\eta_3$ in \eqref{act21} should be replaced by
\be{repl-eta1}
f(\bar\eta_{2},\bar\eta_{3})f(\bar\eta_{3},\bar\eta_{1})=
\frac{f(\bar\eta_{2},\bar\eta_{4})f(\bar\eta_{2},\bar\eta_{5})f(\bar\eta_{5},\bar\eta_{1})f(\bar\eta_{4},\bar\eta_{1})}
{f(\bar\eta_{2},\bar y)f(\bar y,\bar\eta_{1})}.
\ee
Similarly
\be{repl-xi1}
\frac{f(\bar\xi_{2},\bar\xi_{1})}{f(\bar\xi_{2},\bar\eta_{1})}=
\frac{f(\bar\xi_{3},\bar\xi_{1})f(\bar\xi_{4},\bar\xi_{1})f(\bar\xi_{5},\bar\xi_{1})f(\bar y,\bar\eta_{1})}
{f(\bar\xi_{3},\bar\eta_{1})f(\bar\xi_{4},\bar\eta_{1})f(\bar\xi_{5},\bar\eta_{1})f(\bar y,\bar\xi_{1})}.
\ee
Then using \eqref {A-act31-1} we obtain
 \begin{multline}\label{act2131}
S_1=(-q)^{a-k}\lambda_2(\bar x)\lambda_2(\bar y)\,\sum
 r_1(\bar\eta_{1})r_1(\bar\eta_{5})r_3(\bar\xi_{3})\;
\Izerr_{a-k}(q^{-2}\bar x|\bar\eta_{2})
  \Izerl_{a-k}(\bar\eta_{1}|q^2\bar\xi_{1})\num
  \times\Izerl_{a-k}(\bar\xi_{1}|q^2\bar x)
\Izerr_{k}(q^{-2}\bar\eta_{4}|\bar\xi_{3})\Izerl_{k}(\bar\eta_{5}|q^{2}\bar\xi_{4})
\Izerr_{k}(q^{-2}\bar y|\bar\eta_{4})\Izerl_{k}(\bar\xi_{4}|q^{2}\bar y)\num
 \times f(\bar\eta_{2},\bar\eta_{4})f(\bar\eta_{2},\bar\eta_{1}) f(\bar\eta_{2},\bar\eta_{5})
 f(\bar\eta_{4},\bar\eta_{5})f(\bar\eta_{4},\bar\eta_{1}) f(\bar\eta_{5},\bar\eta_{1})
  \num
 \times  \frac{f(\bar\xi_{3},\bar\xi_{4})f(\bar\xi_{3},\bar\xi_{1})f(\bar\xi_{3},\bar\xi_{5})
 f(\bar\xi_{5},\bar\xi_{4})f(\bar\xi_{5},\bar\xi_{1}) f(\bar\xi_{4},\bar\xi_{1})}
   { f(\bar\xi_{3},\bar\eta_{1})f(\bar\xi_{3},\bar\eta_{5})
    f(\bar\xi_{5},\bar\eta_{1})f(\bar\xi_{5},\bar\eta_{5}) f(\bar\xi_{4},\bar\eta_{1})
   f(\bar y,\bar\xi_{1})f(\bar\eta_{2},\bar y)}
 \,\mathbb{B}^{0,b-k}(\emptyset;\bar\xi_{5}).
 \end{multline}
Here $\eta=\{\bar u,\bar x,\bar y\}$ and $\xi=\{\bar v,\bar x, \bar y\}$. The sum is taken over partitions $\bar\eta\Rightarrow\{\bar\eta_{1},\bar\eta_{2},\bar\eta_{3},\bar\eta_{4},\bar\eta_{5}\}$ and
$\bar\xi\Rightarrow\{\bar\xi_{1},\bar\xi_{2},\bar\xi_{3},\bar\xi_{4},\bar\xi_{5}\}$.

The expression \eqref{act2131} can be slightly simplified. First of all we set
$\{\bar\eta_1,\bar\eta_5\}=\bar\eta_6$  and $\{\bar\xi_1,\bar\xi_4\}=\bar\xi_6$. Then the equation
\eqref{act2131} takes the form
 \begin{multline}\label{act2131-simp}
S_1=(-q)^{a-k}\lambda_2(\bar x)\lambda_2(\bar y)\sum
 r_1(\bar\eta_{6})r_3(\bar\xi_{3})\;
  \left\{\Izerl_{a-k}(\bar\eta_{1}|q^2\bar\xi_{1})\Izerl_{k}(\bar\eta_{5}|q^{2}\bar\xi_{4})\frac{f(\bar\eta_{5},\bar\eta_{1})}
  {f(\bar\xi_{4},\bar\eta_{1})}\right\}\num
  \times \Izerr_{a-k}(q^{-2}\bar x|\bar\eta_{2})\Izerl_{a-k}(\bar\xi_{1}|q^2\bar x)
\Izerr_{k}(q^{-2}\bar\eta_{4}|\bar\xi_{3})
\Izerr_{k}(q^{-2}\bar y|\bar\eta_{4})\Izerl_{k}(\bar\xi_{4}|q^{2}\bar y)\num
 \times  \frac{f(\bar\eta_{2},\bar\eta_{4}) f(\bar\eta_{2},\bar\eta_{6})
 f(\bar\eta_{4},\bar\eta_{6})
 f(\bar\xi_{3},\bar\xi_{6})f(\bar\xi_{3},\bar\xi_{5})f(\bar\xi_{5},\bar\xi_{6})
 f(\bar\xi_{4},\bar\xi_{1})}
   { f(\bar\xi_{3},\bar\eta_{6})
   f(\bar\xi_{5},\bar\eta_{6})
   f(\bar y,\bar\xi_{1})f(\bar\eta_{2},\bar y)}\mathbb{B}^{0,b-k}(\emptyset;\bar\xi_{5}).
 \end{multline}
Now we can take the sum over the partitions
$\bar\eta_6\Rightarrow\{\bar\eta_1,\bar\eta_5\}$ (see the terms in the braces in \eqref{act2131-simp}):
\begin{multline}\label{e1e5-part}
\sum\Izerl_{a-k}(\bar\eta_{1}|q^2\bar\xi_{1}) \Izerl_{k}(\bar\eta_{5}|q^2\bar\xi_{4})
\;  \frac{f(\bar\eta_{5},\bar\eta_{1})}{f(\bar\xi_{4},\bar\eta_{1})}\\
=(-q)^{-k}\sum\Izerl_{a-k}(\bar\eta_{1}|q^2\bar\xi_{1}) \Izerr_{k}(\bar\xi_{4}|\bar\eta_{5})
\;  \frac{f(\bar\eta_{5},\bar\eta_{1})}{f(\bar\xi_{4},\bar\eta_{6})}=\Izerl_{a}(\bar\eta_{6}|q^2\bar\xi_{6}).
\end{multline}
Here we first transformed $\Izerl_{k}$ into $\Izerr_{k}$ via \eqref{K-invers} and then applied
\eqref{Sym-Part-old1}. Thus, the equation \eqref{act2131-simp} turns into
 \begin{multline}\label{act2131-1}
S_1=(-q)^{a-k}\lambda_2(\bar x)\lambda_2(\bar y)\sum
 r_1(\bar\eta_{6})r_3(\bar\xi_{3})\;
\Izerr_{a-k}(q^{-2}\bar x|\bar\eta_{2})
 \Izerl_{a}(\bar\eta_{6}|q^2\bar\xi_{6})\num
  \times
\Izerr_{k}(q^{-2}\bar\eta_{4}|\bar\xi_{3})
\Izerr_{k}(q^{-2}\bar y|\bar\eta_{4}) \left\{ \Izerl_{a-k}(\bar\xi_{1}|q^2\bar x)\Izerl_{k}(\bar\xi_{4}|q^{2}\bar y)
\frac{f(\bar\xi_{4},\bar\xi_{1})}{f(\bar y,\bar\xi_{1})}\right\}\num
 \times  \frac{f(\bar\eta_{2},\bar\eta_{4}) f(\bar\eta_{2},\bar\eta_{6})
 f(\bar\eta_{4},\bar\eta_{6})f(\bar\xi_{3},\bar\xi_{6})f(\bar\xi_{3},\bar\xi_{5})f(\bar\xi_{5},\bar\xi_{6})} { f(\bar\xi_{3},\bar\eta_{6})
   f(\bar\xi_{5},\bar\eta_{6}) f(\bar\eta_{2},\bar y)}\mathbb{B}^{0,b-k}(\emptyset;\bar\xi_{5}).
 \end{multline}
Now we can take the sum over partitions $\bar\xi_{6}\Rightarrow\{\bar\xi_{1},\bar\xi_{4}\}$ (see the terms in the braces in \eqref{act2131-1}):
\begin{multline}\label{x1x4-part}
\sum\Izerl_{a-k}(\bar\xi_{1}|q^2\bar x) \Izerl_{k}(\bar\xi_{4}|q^2\bar y)\;
\frac{f(\bar\xi_{4},\bar\xi_{1})}{f(\bar y,\bar\xi_{1})}\\
=(-q)^{-k}\sum\Izerl_{a-k}(\bar\xi_{1}|q^2\bar x) \Izerr_{k}(\bar y|\bar\xi_{4})\;
\frac{f(\bar\xi_{4},\bar\xi_{1})}{f(\bar y,\bar\xi_{6})}=\Izerl_{a}(\bar\xi_{6}|\{q^2\bar x,q^2\}\bar y).
\end{multline}
We again transformed $\Izerl_{k}$ into $\Izerr_{k}$ via \eqref{K-invers} and  applied
\eqref{Sym-Part-old1}. Thus, we finally arrive at
 \begin{multline}\label{act2131-2}
S_1=(-q)^{a-k}\lambda_2(\bar x)\lambda_2(\bar y)\sum
 r_1(\bar\eta_{6})r_3(\bar\xi_{3})\;\Izerl_{a}(\bar\xi_{6}|\{q^2\bar x,q^2\bar y\})
 \Izerl_{a}(\bar\eta_{6}|q^2\bar\xi_{6})\Izerr_{a-k}(q^{-2}\bar x|\bar\eta_{2})\num
  \times \Izerr_{k}(q^{-2}\bar\eta_{4}|\bar\xi_{3})
\Izerr_{k}(q^{-2}\bar y|\bar\eta_{4})\;
  \frac{f(\bar\eta_{2},\bar\eta_{4})f(\bar\eta_{2},\bar\eta_{6})f(\bar\eta_{4},\bar\eta_{6})
  f(\bar\xi_{3},\bar\xi_{5})f(\bar\xi_{5},\bar\xi_{6})
f(\bar\xi_{3},\bar\xi_{6})}
   {f(\bar\xi_{3},\bar\eta_{6})f(\bar\xi_{5},\bar\eta_{6})
   f(\bar\eta_{2},\bar y)}\mathbb{B}^{0,b-k}(\emptyset;\bar\xi_{5}).
 \end{multline}
In this formula $\eta=\{\bar u,\bar x,\bar y\}$ and $\xi=\{\bar v,\bar x, \bar y\}$. The sum is taken over partitions $\bar\eta\Rightarrow\{\bar\eta_{2},\bar\eta_{4},\bar\eta_{6}\}$ and
$\bar\xi\Rightarrow\{\bar\xi_{3},\bar\xi_{5},\bar\xi_{6}\}$. The partitions are independent except their cardinalities
that are fixed by the subscripts of the Izergin determinants and the superscript of Bethe vector.

\subsection{Successive action of $T_{32}(\bar z)T_{31}(\bar y)T_{21}(\bar x)$\label{Astep-2}}

For the computation of the successive action $T_{32}(\bar z)T_{31}(\bar y)T_{21}(\bar x)$ on the
Bethe vector we should act with the product $T_{32}(\bar z)$ on $S_1$. Recall that $\#\bar z=b-k$, therefore we
can use \eqref{A-act32-1}. Then the action of $T_{32}(\bar z)$ gives us an additional sum over partitions $\{\bar z,\bar\xi_{5}\}\Rightarrow\{\bar\xi_{7},\bar\xi_{8}\}$. We have
 \begin{multline}\label{S2-1}
T_{32}(\bar z)S_1=(-q)^{a-k}\lambda_2(\bar x)\lambda_2(\bar y)\lambda_2(\bar z)\sum
 r_1(\bar\eta_{6})r_3(\bar\xi_{3})r_3(\bar\xi_{7})\;\Izerl_{a}(\bar\xi_{6}|\{q^2\bar x,q^2\bar y\})
 \Izerl_{a}(\bar\eta_{6}|q^2\bar\xi_{6})\num
  \times \Izerr_{a-k}(q^{-2}\bar x|\bar\eta_{2})\Izerr_{k}(q^{-2}\bar\eta_{4}|\bar\xi_{3})
\Izerr_{k}(q^{-2}\bar y|\bar\eta_{4})
\Izerr_{b-k}(q^{-2}\bar z|\bar\xi_{7})\Izerl_{b-k}(\bar\xi_{8}|q^2\bar z)\num
\times
  \frac{f(\bar\eta_{2},\bar\eta_{4})f(\bar\eta_{2},\bar\eta_{6})f(\bar\eta_{4},\bar\eta_{6})
  f(\bar\xi_{3},\bar\xi_{7})f(\bar\xi_{3},\bar\xi_{8})f(\bar\xi_{7},\bar\xi_{8}) f(\bar\xi_{8},\bar\xi_{6})
f(\bar\xi_{3},\bar\xi_{6})f(\bar\xi_{7},\bar\xi_{6})f(\bar z,\bar\eta_{6})}
   {f(\bar\xi_{3},\bar\eta_{6})f(\bar\xi_{7},\bar\eta_{6})f(\bar\xi_{3},\bar z)f(\bar z,\bar\xi_{6})
   f(\bar\xi_{8},\bar\eta_{6})   f(\bar\eta_{2},\bar y)}
 \,|0\rangle.
 \end{multline}
In this formula we still have $\eta=\{\bar u,\bar x,\bar y\}$, but $\xi=\{\bar v,\bar x, \bar y, \bar z\}$. The sum is taken over partitions $\bar\eta\Rightarrow\{\bar\eta_{2},\bar\eta_{4},\bar\eta_{6}\}$ and
$\bar\xi\Rightarrow\{\bar\xi_{3},\bar\xi_{6},\bar\xi_{7},\bar\xi_{8}\}$.

Again partial summation over partitions is possible.
Setting $\{\bar\xi_{3},\bar\xi_{7}\}=\bar\xi_{9}$  we obtain
 \begin{multline}\label{S2-2}
T_{32}(\bar z)S_1=(-q)^{a-k}\lambda_2(\bar x)\lambda_2(\bar y)\lambda_2(\bar z)\sum
 r_1(\bar\eta_{6})r_3(\bar\xi_{9})\;\Izerl_{a}(\bar\xi_{6}|\{q^2\bar x,q^2\bar y\})
 \Izerl_{a}(\bar\eta_{6}|q^2\bar\xi_{6})\num
  \times \Izerr_{a-k}(q^{-2}\bar x|\bar\eta_{2})
\Izerr_{k}(q^{-2}\bar y|\bar\eta_{4})
\Izerl_{b-k}(\bar\xi_{8}|q^2\bar z)
\left\{\Izerr_{k}(q^{-2}\bar\eta_{4}|\bar\xi_{3})\Izerr_{b-k}(q^{-2}\bar z|\bar\xi_{7})
\frac{f(\bar\xi_{3},\bar\xi_{7})}{f(\bar\xi_{3},\bar z)}\right\}\num
\times
  \frac{f(\bar\eta_{2},\bar\eta_{4})f(\bar\eta_{2},\bar\eta_{6})f(\bar\eta_{4},\bar\eta_{6})
  f(\bar\xi_{9},\bar\xi_{8}) f(\bar\xi_{8},\bar\xi_{6})
 f(\bar\xi_{9},\bar\xi_{6})f(\bar z,\bar\eta_{6})}
   {f(\bar\xi_{9},\bar\eta_{6})f(\bar z,\bar\xi_{6})
   f(\bar\xi_{8},\bar\eta_{6})   f(\bar\eta_{2},\bar y)}
 \,|0\rangle.
 \end{multline}
In the same manner as before we  take the sum over partitions $\bar\xi_{9}\Rightarrow\{\bar\xi_{3},\bar\xi_{7}\}$ in the braces in
\eqref{S2-2}:
\begin{multline}\label{x2x4-part}
\sum \Izerr_{k}(q^{-2}\bar\eta_{4}|\bar\xi_{3})\Izerr_{b-k}(q^{-2}\bar z|\bar\xi_{7})
\frac{f(\bar\xi_{3},\bar\xi_{7})}{f(\bar\xi_{3},\bar z)}\num
=(-q)^{b-k}\sum \Izerr_{k}(q^{-2}\bar\eta_{4}|\bar\xi_{3})\Izerl_{b-k}(\bar\xi_{7}|\bar z)
\frac{f(\bar\xi_{3},\bar\xi_{7})}{f(\bar\xi_{9},\bar z)}=\Izerr_{b}(\{q^{-2}\bar z,q^{-2}\bar\eta_{4}\}|\bar\xi_{9}).
\end{multline}
Thus, we obtain
 \begin{multline}\label{S2-3}
T_{32}(\bar z)S_1=(-q)^{a-k}\lambda_2(\bar x)\lambda_2(\bar y)\lambda_2(\bar z)\sum
 r_1(\bar\eta_{6})r_3(\bar\xi_{9})\;\Izerl_{a}(\bar\xi_{6}|\{q^2\bar x,q^2\bar y\})
 \Izerl_{a}(\bar\eta_{6}|q^2\bar\xi_{6})\num
  \times \Izerr_{b}(\{q^{-2}\bar z,q^{-2}\bar\eta_{4}\}|\bar\xi_{9})\Izerr_{a-k}(q^{-2}\bar x|\bar\eta_{2})
\Izerr_{k}(q^{-2}\bar y|\bar\eta_{4})\Izerl_{b-k}(\bar\xi_{8}|q^2\bar z)\num
\times
  \frac{f(\bar\eta_{2},\bar\eta_{4})f(\bar\eta_{2},\bar\eta_{6})f(\bar\eta_{4},\bar\eta_{6})
  f(\bar\xi_{9},\bar\xi_{8}) f(\bar\xi_{8},\bar\xi_{6})
 f(\bar\xi_{9},\bar\xi_{6})f(\bar z,\bar\eta_{6})}
   {f(\bar\xi_{9},\bar\eta_{6})f(\bar z,\bar\xi_{6})
   f(\bar\xi_{8},\bar\eta_{6})   f(\bar\eta_{2},\bar y)}
 \,|0\rangle.
 \end{multline}

Finally, we relabel subsets in \eqref{S2-3} as: $\bar\eta_{6}\to\bar\eta_{\rm i}$, $\bar\eta_{2}\to\bar\eta_{\rm ii}$, $\bar\eta_{4}\to\bar\eta_{\rm iii}$
$\bar\xi_{9}\to\bar\xi_{\rm i}$, $\bar\xi_{6}\to\bar\xi_{\rm ii}$, $\bar\xi_{8}\to\bar\xi_{\rm iii}$ and set
$\bar u=\blab$, $\bar v=\bmub$. Then we have
 \begin{multline}\label{S2-res}
T_{32}(\bar z)T_{31}(\bar y)T_{21}(\bar x)\mathbb{B}^{a,b}(\blab;\bmub)=(-q)^{a-k}\lambda_2(\bar x)\lambda_2(\bar y)\lambda_2(\bar z)\sum
 r_1(\bar\eta_{\rm i})r_3(\bar\xi_{\rm i})\Izerl_{a}(\bar\eta_{\rm i}|q^2\bar\xi_{\rm ii})\num
\times \Izerl_{a}(\bar\xi_{\rm ii}|\{q^2\bar x,q^2\bar y\})
 \Izerr_{b}(\{q^{-2}\bar z,q^{-2}\bar\eta_{\rm iii}\}|\bar\xi_{\rm i})\Izerr_{a-k}(q^{-2}\bar x|\bar\eta_{\rm ii})
\Izerr_{k}(q^{-2}\bar y|\bar\eta_{\rm iii})\Izerl_{b-k}(\bar\xi_{\rm iii}|q^2\bar z)\num
\times
  \frac{f(\bar\eta_{\rm ii},\bar\eta_{\rm iii})f(\bar\eta_{\rm ii},\bar\eta_{\rm i})f(\bar\eta_{\rm iii},\bar\eta_{\rm i})
  f(\bar\xi_{\rm i},\bar\xi_{\rm iii}) f(\bar\xi_{\rm iii},\bar\xi_{\rm ii})
 f(\bar\xi_{\rm i},\bar\xi_{\rm ii})f(\bar z,\bar\eta_{\rm i})}
   {f(\bar\xi_{\rm i},\bar\eta_{\rm i})f(\bar z,\bar\xi_{\rm ii})
   f(\bar\xi_{\rm iii},\bar\eta_{\rm i})   f(\bar\eta_{\rm ii},\bar y)}
 \,|0\rangle.
 \end{multline}
Here
\begin{itemize}
\item
$\{\bar x,\bar y,\blab\}=\bar\eta\Rightarrow\{\bar\eta_{\rm i},\bar\eta_{\rm ii},\bar\eta_{\rm iii}\}$;

\item
$\{\bar x,\bar y,\bar z,\bmub\}=\bar\xi\Rightarrow
\{\bar\xi_{\rm i},\bar\xi_{\rm ii},\bar\xi_{\rm iii}\}$.

\end{itemize}

\section{Scalar product\label{S-ScalP}}
\subsection{Scalar product in terms of the highest coefficients}
The equation \eqref{S2-res} allows us to obtain an expression for the scalar product of Bethe vectors \eqref{Def-SP} in
terms of sums over partitions of Bethe parameters.  For this we take a representation
for the dual Bethe vector \eqref{Or-form}
\be{dual-BV}
\mathbb{C}^{a,b}(\blac;\bmuc) =\sum \frac{(-q)^{k}\Izerl_{k}(\blac_{\rm i}|q^2\bmuc_{\rm i})}
{\as_2(\bmuc_{\rm ii})\as_2(\blac)f(\bmuc_{\rm ii},\blac)}
f(\bmuc_{\rm ii},\bmuc_{\rm i})f(\blac_{\rm ii},\blac_{\rm i})\,
\langle0|T_{32}(\bmuc_{\rm ii})T_{31}(\blac_{\rm i})T_{21}(\blac_{\rm ii}).
\ee
Recall that here the sum is taken with respect to partitions $\blac\Rightarrow\{\blac_{\rm i},\blac_{\rm ii}\}$ and
$\bmuc\Rightarrow\{\bmuc_{\rm i},\bmuc_{\rm ii}\}$ with $\#\blac_{\rm i}=\#\bmuc_{\rm i}=k$ and $k=0,1,\dots,\min(a,b)$.
Thus, in order to calculate \eqref{Def-SP} we should take \eqref{dual-BV} and combine  it with \eqref{S2-res} setting there $\bar y=\blac_{\rm i}$, $\bar x=\blac_{\rm ii}$, and $\bar z=\bmuc_{\rm ii}$.
Then
 \begin{multline}\label{SP-init}
S_{a,b}=
(-q)^{a}\sum
 r_1(\bar\eta_{\rm i})r_3(\bar\xi_{\rm i})\;\Izerl_{a}(\bar\xi_{\rm ii}|q^2\blac) \Izerl_{a}(\bar\eta_{\rm i}|q^2\bar\xi_{\rm ii})
 \Izerr_{b}(\{q^{-2}\bmuc_{\rm ii},q^{-2}\bar\eta_{\rm iii}\}|\bar\xi_{\rm i})\num
\times
\Izerl_{b-k}(\bar\xi_{\rm iii}|q^2\bmuc_{\rm ii})
\left\{\Izerr_{a-k}(q^{-2}\blac_{\rm ii}|\bar\eta_{\rm ii})
\Izerr_{k}(q^{-2}\blac_{\rm i}|\bar\eta_{\rm iii})
\Izerl_{k}(\blac_{\rm i}|q^2\bmuc_{\rm i})\frac{f(\blac_{\rm ii},\blac_{\rm i})}{f(\bar\eta_{\rm ii},\blac_{\rm i})}\right\}\num
\times
  \frac{f(\bar\eta_{\rm ii},\bar\eta_{\rm iii})f(\bar\eta_{\rm ii},\bar\eta_{\rm i})f(\bar\eta_{\rm iii},\bar\eta_{\rm i})
  f(\bar\xi_{\rm i},\bar\xi_{\rm iii}) f(\bar\xi_{\rm iii},\bar\xi_{\rm ii})
 f(\bar\xi_{\rm i},\bar\xi_{\rm ii})f(\bmuc_{\rm ii},\bar\eta_{\rm i})f(\bmuc_{\rm ii},\bmuc_{\rm i})}
   {f(\bar\xi_{\rm i},\bar\eta_{\rm i})f(\bmuc_{\rm ii},\bar\xi_{\rm ii})
   f(\bar\xi_{\rm iii},\bar\eta_{\rm i})  f(\bmuc_{\rm ii},\blac)}.
 \end{multline}
Recall that here $\{\blac,\blab\}=\bar\eta\Rightarrow\{\bar\eta_{\rm i},\bar\eta_{\rm ii},\bar\eta_{\rm iii}\}$ and
$\{\blac,\bmuc_{\rm ii},\bmub\}=\bar\xi\Rightarrow
\{\bar\xi_{\rm i},\bar\xi_{\rm ii},\bar\xi_{\rm iii}\}$.  For shortness we have also omitted the arguments of
$S_{a,b}(\blac;\bmuc|\blab;\bmub)$ in the l.h.s. of \eqref{SP-init}.

Now the expression \eqref{SP-init} can be simplified. In particular, one can apply \eqref{K-invers}
and  \eqref{Al-RHC-IHC-twin1} to the terms in braces in order to take the sum over
partitions $\blac\Rightarrow\{\blac_{\rm i},\blac_{\rm ii}\}$:
\begin{multline}\label{sum-RHC1}
\sum
\Izerr_{k}(q^{-2}\blac_{\rm i}|\bar\eta_{\rm iii})\Izerl_{k}(\blac_{\rm i}|q^2\bmuc_{\rm i})\Izerr_{a-k}(q^{-2}\blac_{\rm ii}|\bar\eta_{\rm ii})
\frac{f(\blac_{\rm ii},\blac_{\rm i})}{f(\bar\eta_{\rm ii},\blac_{\rm i})}\num
=(-q)^{a-k}\sum
\Izerr_{k}(\blac_{\rm i}|q^{2}\bar\eta_{\rm iii})\Izerl_{k}(\blac_{\rm i}|q^2\bmuc_{\rm i})\Izerl_{a-k}(\bar\eta_{\rm ii}|\blac_{\rm ii})
\frac{f(\blac_{\rm ii},\blac_{\rm i})}{f(\bar\eta_{\rm ii},\blac)}\num
=\frac{(-q)^{a}\RHCl_{a,k}(\blac;\{\bar\eta_{\rm ii},\bar\eta_{\rm iii}\}|\bmuc_{\rm i};q^{-2}\bar\eta_{\rm iii})}
{f(\bar\eta_{\rm ii},\blac)f(\bar\eta_{\rm iii},\blac)f(\bmuc_{\rm i},\blac)}.
\end{multline}
Then we obtain
 \begin{multline}\label{SP-1t}
S_{a,b}=\frac{(-q)^{2a}}{f(\bmuc,\blac)}\sum
 r_1(\bar\eta_{\rm i})r_3(\bar\xi_{\rm i})\;
 \RHCl_{a,k}(\blac;\{\bar\eta_{\rm ii},\bar\eta_{\rm iii}\}|\bmuc_{\rm i};q^{-2}\bar\eta_{\rm iii})\num
\times \Izerr_{b}(\{q^{-2}\bmuc_{\rm ii},q^{-2}\bar\eta_{\rm iii}\}|\bar\xi_{\rm i})\left\{\Izerl_{a}(\bar\xi_{\rm ii}|q^2\blac)
 \Izerl_{a}(\bar\eta_{\rm i}|q^2\bar\xi_{\rm ii}) \Izerl_{b-k}(\bar\xi_{\rm iii}|q^2\bmuc_{\rm ii})
 \frac{f(\bar\xi_{\rm iii},\bar\xi_{\rm ii})}{f(\bmuc_{\rm ii},\bar\xi_{\rm ii}) f(\bar\xi_{\rm iii},\bar\eta_{\rm i})}\right\}
\num
\times
  \frac{f(\bar\eta_{\rm ii},\bar\eta_{\rm iii})f(\bar\eta_{\rm ii},\bar\eta_{\rm i})f(\bar\eta_{\rm iii},\bar\eta_{\rm i})
  f(\bar\xi_{\rm i},\bar\xi_{\rm iii})
 f(\bar\xi_{\rm i},\bar\xi_{\rm ii})f(\bmuc_{\rm ii},\bar\eta_{\rm i})f(\bmuc_{\rm ii},\bmuc_{\rm i})}
   {f(\bar\xi_{\rm i},\bar\eta_{\rm i})  f(\bar\eta_{\rm ii},\blac)f(\bar\eta_{\rm iii},\blac)}.
 \end{multline}

Similarly we can take the sum of the terms in braces of \eqref{SP-1t} over partitions $\bar\xi_{\rm ii}$ and $\bar\xi_{\rm iii}$.
Indeed, due to \eqref{K-invers} and \eqref{Al-RHC-IHC-twin1} we have
\begin{multline}\label{sum-RHC2}
\sum\Izerl_{a}(\bar\xi_{\rm ii}|q^2\blac) \Izerl_{a}(\bar\eta_{\rm i}|q^2\bar\xi_{\rm ii})
\Izerl_{b-k}(\bar\xi_{\rm iii}|q^2\bmuc_{\rm ii})\frac{f(\bar\xi_{\rm iii},\bar\xi_{\rm ii})}{f(\bmuc_{\rm ii},\bar\xi_{\rm ii})
   f(\bar\xi_{\rm iii},\bar\eta_{\rm i})}\num
=(-q)^{k-b-a}\sum\Izerl_{a}(\bar\xi_{\rm ii}|q^2\blac) \Izerr_{a}(\bar\xi_{\rm ii}|\bar\eta_{\rm i})
\Izerr_{b-k}(\bmuc_{\rm ii}|\bar\xi_{\rm iii})\frac{f(\bar\xi_{\rm iii},\bar\xi_{\rm ii})}{f(\bmuc_{\rm ii},\bar\xi_{0})
   f(\bar\xi_{0},\bar\eta_{\rm i})}\num
=(-q)^{k-b-2a}\frac{\RHCr_{a+b-k,a}(\bar\xi_0;\{\bmuc_{\rm ii},\blac\}|q^{-2}\bar\eta_{\rm i};q^{-2}\blac)}
{f(\bmuc_{\rm ii},\bar\xi_0)f(\blac,\bar\xi_0)},
\end{multline}
where $\bar\xi_{0}=\{\bar\xi_{\rm ii},\bar\xi_{\rm iii}\}$. Thus, we obtain
 \begin{multline}\label{SP-2t}
S_{a,b}=\sum\frac{(-q)^{k-b}}{f(\bmuc,\blac)}
 r_1(\bar\eta_{\rm i})r_3(\bar\xi_{\rm i})\;\Izerr_{b}(\{q^{-2}\bmuc_{\rm ii},q^{-2}\bar\eta_{\rm iii}\}|\bar\xi_{\rm i})
 \;\RHCl_{a,k}(\blac;\bar\eta_{0}|\bmuc_{\rm i};q^{-2}\bar\eta_{\rm iii})\num
\times \RHCr_{a+b-k,a}(\bar\xi_0;\{\bmuc_{\rm ii},\blac\}|q^{-2}\bar\eta_{\rm i};q^{-2}\blac)\;
  \frac{f(\bar\eta_{\rm ii},\bar\eta_{\rm iii})f(\bar\eta_{0},\bar\eta_{\rm i})
  f(\bar\xi_{\rm i},\bar\xi_{0})f(\bmuc_{\rm ii},\bar\eta_{\rm i})f(\bmuc_{\rm ii},\bmuc_{\rm i})}
   {f(\bar\xi_{\rm i},\bar\eta_{\rm i}) f(\bar\eta_{0},\blac)f(\bmuc_{\rm ii},\bar\xi_0)f(\blac,\bar\xi_0)},
 \end{multline}
where $\bar\eta_{0}=\{\bar\eta_{\rm ii},\bar\eta_{\rm iii}\}$.

{\sl Remark.} Strictly speaking, one should understand the sets $\bar\eta$ and $\bar\xi$ in the equation \eqref{SP-2t} as
\be{eps}
\begin{array}{l}
\bar\eta=\{\blab+\epsilon,\blac+\epsilon\},\\
\bar\xi=\{\blac+\epsilon,\bmub+\epsilon,\bmuc_{\rm ii}+\epsilon\},
\end{array}
\qquad\text{at}\quad \epsilon\to 0.
\ee
The matter is that individual
factors in \eqref{SP-2t} may have singularities, if we set $\epsilon=0$. For instance, the highest
coefficient $\RHCl_{a,k}(\blac;\bar\eta_{0}|\bmuc_{\rm ii};q^{-2}\bar\eta_{\rm iii})$ is singular if $\bar\eta_{0}\cap\blac
\ne\emptyset$ (see \eqref{2-red-1}). These singularities, of course, are compensated by the product $f^{-1}(\bar\eta_{0},\blac)$,
but for evaluating the limit we should have $\epsilon\ne 0$. In order to lighten the formulas we do
not write this auxiliary parameter $\epsilon$ explicitly, but one has to keep it in mind when doing the calculations.

The equation \eqref{SP-2t} already gives the representation for the scalar product of Bethe vectors in term of the sum
over partitions of Bethe parameters. However it is not convenient for further applications. In particular, it
contains many terms which actually cancel each other. Below we simplify \eqref{SP-2t} making several reductions of
the highest coefficients.

\subsection{The first reduction of the highest coefficients \label{S-RedHC1}}

The first simplification of \eqref{SP-2t} is based on the following

\begin{prop}\label{P-uc}
The subset $\bar\xi_{\rm i}$ in \eqref{SP-2t} does not contain the elements from the set $\blac$,
that is $\bar\xi_{\rm i}\cap \blac=\emptyset$.
\end{prop}

{\sl Proof.} The representation for the scalar product \eqref{SP-2t} can be written in the following form
\be{Sp-short1}
S_{a,b}\equiv\mathbb{C}^{a,b}(\blac;\bmuc)\mathbb{B}^{a,b}(\blab;\bmub)=
\sum r_1(\bar\eta_{\rm i})r_3(\bar\xi_{\rm i})\; W_{\text{part}}(\bar\eta;\bar\xi;q),
\ee
where $W_{\text{part}}(\bar\eta;\bar\xi;q)$ is a rational function depending on the partitions. For the moment
its explicit form is not important, however we have shown explicitly that $W_{\text{part}}$ depends on the parameter $q$.
The sum is taken over partitions of the set
$\bar\eta=\{\blab,\blac\}$ and of the  set $\bar\xi=\{\blac,\bmub,\bmuc_{\rm ii}\}$.

We can apply the isomorphism $\varphi$ to the equation \eqref{Sp-short1}. Due to \eqref{phi-F} in the r.h.s. one should
simply make the replacement $r_1\leftrightarrow r_3$:
\be{Sp-short2}
\varphi(S_{a,b})=
\sum r_3(\bar\eta_{\rm i})r_1(\bar\xi_{\rm i})\; W_{\text{part}}(\bar\eta;\bar\xi;q).
\ee
On the other hand due to \eqref{phi-BV} we have in the l.h.s.
\be{Sp-short3}
\varphi(S_{a,b})= \varphi\bigl((\mathbb{C}^{a,b}(\blac;\bmuc)\bigr)
\varphi\bigl(\mathbb{B}^{a,b}(\blab;\bmub)\bigr)=\mathbb{C}_{q^{-1}}^{b,a}(\bmuc;\blac)
\mathbb{B}_{q^{-1}}^{b,a}(\bmub;\blab).
\ee
Calculating the scalar product in \eqref{Sp-short3} via \eqref{Sp-short1} we obtain
\be{Sp-short4}
\varphi(S_{a,b})=
\sum r_1(\bar\eta'_{\rm i})r_3(\bar\xi'_{\rm i})\; W_{\text{part}}(\bar\eta';\bar\xi';q^{-1}),
\ee
where now the sum is taken over partitions of the set
$\bar\eta'=\{\bmub,\bmuc\}$ and the set $\bar\xi'=\{\bmuc,\blab,\blac_{\rm ii}\}$. Thus, we arrive at
\be{Sp-short5}
\sum r_1(\bar\eta'_{\rm i})r_3(\bar\xi'_{\rm i})\; W_{\text{part}}(\bar\eta';\bar\xi';q^{-1})=
\sum r_3(\bar\eta_{\rm i})r_1(\bar\xi_{\rm i})\; W_{\text{part}}(\bar\eta;\bar\xi;q).
\ee
Since $r_1$ and $r_3$ are free functional parameters we conclude that for an arbitrary subset
$\bar\eta'_{\rm i}\subset\bar\eta'$ there exists a subset $\bar\xi_{\rm i}\subset\bar\xi$ such that
$\bar\eta'_{\rm i}=\bar\xi_{\rm i}$
and vice versa. But $\bar\eta'_{\rm i}\subset\{\bmub,\bmuc\}$, hence, $\bar\xi_{\rm i}\subset\{\bmub,\bmuc\}$, and thus
$\bar\xi_{\rm i}\cap\blac=\emptyset$.\qed

Due to Proposition~\ref{P-uc} if $\bar\xi_{\rm i}\cap \blac\ne\emptyset$, then the corresponding term in the
sum \eqref{SP-2t} vanishes. Hence, $\blac\subset\bar\xi_0$, and we can set
$\bar\xi_0=\{\bar\xi_{\rm ii},\blac\}$ and $\{\bar\xi_{\rm i},\bar\xi_{\rm ii}\}=\bar\xi=\{\bmub,\bmuc_{\rm ii}\}$. Then we obtain
a possibility to simplify the highest coefficient $\RHCr_{a+b-k,a}$ via \eqref{2-red-1} and \eqref{Z-invers}
\begin{multline}\label{Reduct-1}
\frac1{f(\bar\xi_0,\blac)}\RHCr_{a+b-k,a}(\{\bar\xi_{\rm ii},\blac\};\{\bmuc_{\rm ii},\blac\}|q^{-2}\bar\eta_{\rm i};q^{-2}\blac)
=\frac{f(\bmuc_{\rm ii},\blac)}{f(\blac,\bar\eta_{\rm i})}
\RHCr_{b-k,a}(\bar\xi_{\rm ii};\bmuc_{\rm ii}|q^{-2}\bar\eta_{\rm i};q^{-2}\blac)\num
=\frac{
\RHCr_{a,b-k}(\bar\eta_{\rm i};\blac|\bar\xi_{\rm ii};\bmuc_{\rm ii})}
{f(\blac,\bar\eta_{\rm i})f(\bar\xi_{\rm ii},\bar\eta_{\rm i})}.
\end{multline}
Thus, the equation \eqref{SP-2t} turns into
 \begin{multline}\label{SP-1-red}
S_{a,b}=\sum \frac{(-q)^{k-b}}{f(\bmuc,\blac)}
 r_1(\bar\eta_{\rm i})r_3(\bar\xi_{\rm i})\;\Izerr_{b}(\{q^{-2}\bmuc_{\rm ii},q^{-2}\bar\eta_{\rm iii}\}|\bar\xi_{\rm i})
 \;\RHCl_{a,k}(\blac;\bar\eta_{0}|\bmuc_{\rm i};q^{-2}\bar\eta_{\rm iii})\num
\times \RHCr_{a,b-k}(\bar\eta_{\rm i};\blac|\bar\xi_{\rm ii};\bmuc_{\rm ii})\;
  \frac{f(\bar\eta_{\rm ii},\bar\eta_{\rm iii})f(\bar\eta_{0},\bar\eta_{\rm i})
  f(\bar\xi_{\rm i},\bar\xi_{\rm ii})f(\bar\xi_{\rm i},\blac)f(\bmuc_{\rm ii},\bmuc_{\rm i})}
   {f(\bmub,\bar\eta_{\rm i}) f(\bar\eta_{0},\blac)f(\bmuc_{\rm ii},\blac)f(\bmuc_{\rm ii},\bar\xi_{\rm ii})f(\blac,\bar\eta_{\rm i})},
 \end{multline}
where we have used $\{\bar\xi_{\rm i},\bar\xi_{\rm ii}\}=\{\bmub,\bmuc_{\rm ii}\}$.

\subsection{The second reduction}

For further reductions we should specify the subsets $\bar\eta_k$ and $\bar\xi_k$ in terms of the
original variables. Recall that the sum in \eqref{SP-1-red} is taken over partitions:

\begin{itemize}
\item $\{\bmub,\bmuc_{\rm ii}\}\Rightarrow\{\bar\xi_{\rm i},\bar\xi_{\rm ii}\}$;

\item $\{\blac,\blab\}\Rightarrow\{\bar\eta_{\rm i},\bar\eta_{0}\}$ and then
$\bar\eta_{0}\Rightarrow\{\bar\eta_{\rm ii},\bar\eta_{\rm iii}\}$.
\end{itemize}

We set
\be{set-u}
\begin{aligned}
\bar\eta_{\rm i}&=\{\blab_{\so},\blac_{\st}\},\\
\bar\eta_{0}&=\{\blab_{\st},\blac_{\so}\},
\end{aligned}
\ee
with $\#\blac_{\so}=\#\blab_{\so}=n$, $n=0,1\dots,a$.  Note that we do not specify subsets $\bar\eta_{\rm ii}$ and $\bar\eta_{\rm iii}$,
however $\{\bar\eta_{\rm ii},\bar\eta_{\rm iii}\}=\{\blab_{\st},\blac_{\so}\}$.

The partitions of the sets $\bmub$ and $\bmuc$ are more
sophisticated.

We start with $\bmuc$, that appears  in \eqref{SP-1-red}
as  $\bmuc\Rightarrow \{\bmuc_{\rm i},\bmuc_{\rm ii}\}$ with $k=\#\bmuc_{\rm i}$.
First of all we divide the set $\bmuc_{\rm ii}$ into additional subsets $\bmuc_{\rm ii}\Rightarrow\{\bmuc_{\st},\bmuc_{0}\}$
with $\#\bmuc_{\st}=b-m$, $m=0,1,\dots,b-k$. Then we define the subset $\bmuc_{\so}$ such that
$\{\bmuc_{\so},\bmuc_{\st}\}=\bmuc$. Evidently $\#\bmuc_{\so}=m$ and $\bmuc_{\so}=\{\bmuc_{\rm i},\bmuc_{0}\}$.

Finally we
divide the set $\bmub$ as $\bmub\Rightarrow\{\bmub_{\so},\bmub_{\st}\}$ with $\#\bmub_{\so}=m$ and set
\be{set-v1}
\begin{aligned}
&\bar\xi_{\rm i}=\{\bmub_{\so},\bmuc_{\st}\},\\
&\bar\xi_{\rm ii}=\{\bmub_{\st},\bmuc_{0}\}.
\end{aligned}
\ee
Hereby
\be{set-v2}
\begin{aligned}
&\bmuc_{\rm ii}=\{\bmuc_{\st}, \bmuc_{0}\},\\
&\bmuc_{\rm i}=\bmuc_{\so}\setminus\bmuc_{0},
\end{aligned}\qquad
\begin{aligned}
&\#\bmuc_{\so}=\#\bmub_{\so}=m,\\
&\#\bmuc_{\st}=\#\bmub_{\st}=b-m,
\end{aligned}\qquad
\begin{aligned}
&\#\bmuc_{\rm i}=k,\\
&\#\bmuc_{0}=m-k.
\end{aligned}
\ee

Now we can apply \eqref{2-red-1} to the highest coefficient $\RHCl_{a,k}$:
\begin{equation}\label{Reduct-2}
\frac1{f(\bar\eta_{0},\blac)}\RHCl_{a,k}(\blac;\bar\eta_{0}|\bmuc_{\rm i};q^{-2}\bar\eta_{\rm iii})=
\frac{f(\bmuc_{\rm i},\blac_{\so})}{f(\blab_{\st},\blac_{\st})}\RHCl_{a-n,k}(\blac_{\st};\blab_{\st}|\bmuc_{\rm i};q^{-2}\bar\eta_{\rm iii}).
\end{equation}

For simplification of $\RHCr_{a,b-k}$ we use successively \eqref{2-red-1} and then \eqref{2-red-2}
\begin{equation}\label{Reduct-3}
\frac{\RHCr_{a,b-k}(\bar\eta_{\rm i};\blac|\bar\xi_{\rm ii};\bmuc_{\rm ii})}{f(\blac,\bar\eta_{\rm i})f(\bmuc_{\rm ii},\bar\xi_{\rm ii})}
=\frac{f(\bmub_{\st},\blac_{\st})f(\bmuc_{0},\blac)}{f(\blac_{\so},\blab_{\so})f(\bmuc_{\st},\bmub_{\st})}
\;\RHCr_{n,b-m}(\blab_{\so};\blac_{\so}|\bmub_{\st};\bmuc_{\st}).
\end{equation}

Finally, due to \eqref{K-red} we have
\begin{equation}\label{Reduct-4}
\Izerr_{b}(\{q^{-2}\bmuc_{\rm ii},q^{-2}\bar\eta_{\rm iii}\}|\bar\xi_{\rm i})=(-q)^{b-m}
\Izerr_{m}(\{q^{-2}\bmuc_{0},q^{-2}\bar\eta_{\rm iii}\}|\bmub_{\so}).
\end{equation}

Now we should substitute \eqref{Reduct-2}--\eqref{Reduct-4} into \eqref{SP-1-red}. After some elementary but
rather exhaus\-ting algebra we obtain
 \begin{multline}\label{SP-3t}
S_{a,b}=f^{-1}(\bmuc,\blac)f^{-1}(\bmub,\blab)\sum
 r_1(\blab_{\so})r_1(\blac_{\st})r_3(\bmub_{\so})r_3(\bmuc_{\st})\;\RHCr_{n,b-m}(\blab_{\so};\blac_{\so}|\bmub_{\st};\bmuc_{\st})\num
\times  (-q)^{k-m}\Izerr_{m}(q^{-2}\bmuc_{0},q^{-2}\bar\eta_{\rm iii}|\bmub_{\so}) \RHCl_{a-n,k}(\blac_{\st};\blab_{\st}|\bmuc_{\rm i};q^{-2}\bar\eta_{\rm iii})
f(\bar\eta_{\rm ii},\bar\eta_{\rm iii})f(\bmuc_{0},\bmuc_{\rm i})f(\bmub_{\so},\bmuc_{0})f(\bmuc_{\rm i},\blac_{\so})\num
 \times
 f(\blab_{\st},\blab_{\so}) f(\blac_{\so},\blac_{\st})f(\bmub_{\so},\bmub_{\st})f(\bmuc_{\st},\bmuc_{\so})f(\bmub_{\so},\blac_{\so})
 f(\bmub,\blab_{\st}).
 \end{multline}
Recall that here $\{\bar\eta_{\rm ii},\bar\eta_{\rm iii}\}=\{\blab_{\st},\blac_{\so}\}$.

Thus, the representation for the scalar product $S_{a,b}$ has the form \eqref{scal-res}, where
the function $W_{\text{part}}$ has the
following representation:
\begin{multline}\label{Wpart}
W_{\text{part}}\begin{pmatrix}\blac_{\st},\blab_{\st},&\blac_{\so},\blab_{\so}\\
\bmuc_{\so},\bmub_{\so},&\bmuc_{\st},\bmub_{\st}\end{pmatrix}=
\RHCr_{n,b-m}(\blab_{\so};\blac_{\so}|\bmub_{\st};\bmuc_{\st})\\
\times f(\blab_{\st},\blab_{\so}) f(\blac_{\so},\blac_{\st})f(\bmub_{\so},\bmub_{\st})f(\bmuc_{\st},\bmuc_{\so})f(\bmub_{\so},\blac_{\so})
 f(\bmuc_{\so},\blac_{\so})f(\bmub,\blab_{\st})\;\widetilde W.
  \end{multline}
The factor $\widetilde W$ is equal to the sum over additional partitions $\bmuc_{\so}\Rightarrow\{\bmuc_{0},\bmuc_{\rm i}\}$ and $\{\blab_{\st},\blac_{\so}\}\Rightarrow\{\bar\eta_{\rm ii},\bar\eta_{\rm iii}\}$:
\begin{multline}\label{tW}
\widetilde W=\sum  (-q)^{k-m}\Izerr_{m}(q^{-2}\bmuc_{0},q^{-2}\bar\eta_{\rm iii}|\bmub_{\so}) \RHCl_{a-n,k}(\blac_{\st};\blab_{\st}|\bmuc_{\rm i};q^{-2}\bar\eta_{\rm iii})\\
\times f(\bar\eta_{\rm ii},\bar\eta_{\rm iii})f(\bmuc_{0},\bmuc_{\rm i})f(\bmub_{\so},\bmuc_{0})f^{-1}(\bmuc_{0},\blac_{\so}).
\end{multline}
This sum can be explicitly calculated  due to the following identity.

\begin{prop}\label{MainProp1}
Let $a$, $b$, $n$ be non-negative integers and $\bar t$, $\bar x$, $\bar s$, $\bar y$, $\bar z$ be
five sets of generic complex variables with cardinalities $\#\bar t=\#\bar x=a$, $\#\bar s=\#\bar y=b$, and
$\#\bar z=n$. Then
\begin{multline}\label{Sum-F}
f^{-1}(\bar y,\bar\eta)\RHClr_{a,b}(\bar t;\bar x|\bar s;\bar y)=\sum (-q)^{\mp \ell}
\Izerrl_b(\{\bar s_{\so}q^{-2},\bar\eta_{\so}q^{-2}\}|\bar y)
\RHClr_{a,b-k}(\bar t;\bar x|\bar s_{\st};\bar\eta_{\so}q^{-2}) \\
\times f(\bar s_{\so},\bar s_{\st})f(\bar \eta_{\st},\bar \eta_{\so})
f(\bar y,\bar s_{\so})f^{-1}(\bar s_{\so},\bar z).
\end{multline}
Here $\bar\eta$ is a union of two sets: $\bar\eta=\{\bar x,\bar z\}$. The sum is taken
over partitions of the set $\bar s\Rightarrow\{\bar s_{\so},\bar s_{\st}\}$ with $\#\bar s_{\so}=\ell\in[0,\dots,b]$
and the set $\bar \eta\Rightarrow\{\bar\eta_{\so},\bar\eta_{\st}\}$ with $\#\bar\eta_{\so}=b-\ell$.
\end{prop}
This identity was proved in \cite{PakRS13c}. We give another proof in appendix~\ref{A-PP}.
It is easy to see that making
in \eqref{Sum-F} the following change of variables:

\begin{itemize}

\item $\bar\eta_{\so}\to\bar\eta_{\rm iii}$ and  $\bar\eta_{\st}\to\bar\eta_{\rm iii}$;

\item $\bar s_{\so}\to\bmuc_{0}$ and  $\bar s_{\st}\to\bmuc_{\rm i}$;

\item $\bar y=\bmub_{\so}$, $\bar t=\blac_{\st}$, $\bar x=\blab_{\st}$, $\bar z=\blac_{\so}$;

\item $a\to a-n$,  $\ell\to m-k$, $b\to m$.
\end{itemize}
we reproduce the sum over partitions in \eqref{tW}, what gives us

 \begin{equation}\label{Very-Non-triv}
\widetilde W
 =\frac{\RHCl_{a-n,m}(\blac_{\st};\blab_{\st}|\bmuc_{\so};\bmub_{\so})}{f(\bmub_{\so},\blac_{\so})f(\bmuc_{\so},\blab_{\st})} .
 \end{equation}
Substituting this into \eqref{SP-3t} we reproduce \eqref{W-Reshet}. Thus, we have arrived at the trigonometric analog of the formula
for the scalar product of Bethe vectors obtained in \cite{Res86} for the models with $GL(3)$-invariant $R$-matrix.
%
%
%

\section*{Conclusion}

The main goal of this paper was to prove the representation \eqref{scal-res}, \eqref{W-Reshet} for the scalar
product of Bethe vectors in the models with $GL(3)$ trigonometric $R$-matrix. As we have mentioned already, this representation looks very similar to the one obtained in \cite{Res86}. The main difference is that the rational coefficients $W_{\text{part}}$
in \eqref{W-Reshet} depend now on two different highest coefficients, while in the models with $GL(3)$-invariant
$R$-matrix there exist only one highest coefficient.

As we have mentioned in the Introduction, the representation \eqref{scal-res}  is not convenient for
direct applications to the study of correlation functions and form factors of local operators. It is worth
mentioning however, that this representation is obtained for the most general case, when the Bethe
parameters of both vectors are arbitrary complex numbers. Moreover, we considered the vacuum eigenvalues $\lambda_i(u)$
of the operators $T_{ii}(u)$ as free functional parameters. In this case one hardly can hope to simplify
the expression for the scalar product. On the other hand, calculating the correlation functions and form factors
in concrete quantum models, we always deal with a specific representation of $\AQ$ algebra. This fixes the
functions $\lambda_i(u)$. Furthermore, in the case of correlation functions we have to consider particular cases of scalar products, where most of the Bethe parameters satisfy Bethe equations. This gives us a possibility  to calculate at least a part of the sums over partitions in equation \eqref{scal-res}. In this way one can obtain compact determinant representations for some
scalar products and form factors of the monodromy matrix entries. This was done already for the scalar products
in the models with $GL(3)$-invariant $R$-matrix (see \cite{BelPRS12b}, \cite{BelPRS13a}). One can expect that
similar results can be obtained starting from  the representation \eqref{scal-res}, \eqref{W-Reshet}.

\section*{Acknowledgements}

E.R. was supported by ANR Project
DIADEMS (Programme Blanc ANR SIMI1 2010-BLAN-0120-02).
N.A.S. was  supported by the Program of RAS Basic Problems of the Nonlinear Dynamics,
grants RFBR-13-01-12405-ofi-m2.

\appendix

\section{Properties of Izergin determinants\label{A-PID}}

Most of the properties of the left and right Izergin determinants easily follow
directly from their definitions. We give below a list of these properties.
We remind that the  superscript $(l,r)$ on $\Izer$ means that the equality is valid for $\Izerl$ and for $\Izerr$ with appropriate choice of component (first/up or second/down) throughout the equality.

Initial condition:
 \begin{equation}\label{K-init}
\Izerl_{1}(\bar x|\bar y) = x\;g(x,y),\qquad \Izerr_{1}(\bar x|\bar y) = y\;g(x,y).
\end{equation}
Scaling:
 \begin{equation}\label{K-scal}
\Izerlr_{n}(\alpha\bar x|\alpha\bar y) = \Izerlr_{n}(\bar x|\bar y).
\end{equation}
Reduction:
 \begin{equation}\label{K-red}
\Izerlr_{n+1}(\{\bar x, q^{-2}z\}|\{\bar y,z\}) =\Izerlr_{n+1}(\{\bar x, z\}|\{\bar y,q^{2}z\})=-q^{\mp 1}\Izerlr_{n}(\bar x|\bar y).
\end{equation}
Inverse order of arguments:
 \begin{align}\label{K-invers}
  \Izerlr_{n}( q^{-2}\bar x|\bar y)
&= (-q)^{\mp n} f^{-1}(\bar y,\bar x) \Izerrl_{n}(\bar y|\bar x)\,,\\
 \Izerlr_{n;q^{-1}}(\bar x|\bar y)&  =\Izerrl_{n;q}(\bar y|\bar x)\,,\label{K-invers1}
\end{align}
 where $\Izerlr_{n;q^{-1}}$ means $\Izerlr_{n}$ with $q$ replaced by $q^{-1}$.
 We have put an additional index $q^{-1}$ or $q$ in \eqref{K-invers1} to stress this replacement.

Residues in the poles:
 \begin{equation}\label{K-Res}
\Bigl.\Izerlr_{n+1}(\{\bar x,z\}|\{\bar y,z'\})\Bigr|_{z'\to z}= f(z,z')
f(z,\bar y)f(\bar x,z) \Izerlr_{n}(\bar x|\bar y)+{\rm reg},
\end{equation}
where ${\rm reg}$ means the regular part.

The Izergin determinants satisfy also summation identities.

\begin{lemma}\label{main-ident}
Let $\bar\gamma$, $\bar\alpha$ and $\bar\beta$ be three sets of complex variables with $\#\alpha=m_1$,
$\#\beta=m_2$, and $\#\gamma=m_1+m_2$. Then
\begin{equation}\label{Sym-Part-old1}
  \sum
 \Izerlr_{m_1}(\bar\gamma_{\so}|\bar \alpha)\Izerrl_{m_2}(\bar \beta|\bar\gamma_{\st})f(\bar\gamma_{\st},\bar\gamma_{\so})
 = (-q)^{\mp m_1}f(\bar\gamma,\bar \alpha) \Izerrl_{m_1+m_2}(\{\bar \alpha q^{-2},\bar \beta\}|\bar\gamma).
 \end{equation}
The sum is taken with respect to all partitions of the set $\bar\gamma\Rightarrow\{\bar\gamma_{\so},\bar\gamma_{\st}\}$ with $\#\bar\gamma_{\so}=m_1$ and $\#\bar\gamma_{\st}=m_2$.
Due to \eqref{K-invers} the equation \eqref{Sym-Part-old1} can be also written in the form
\begin{equation}\label{Sym-Part-old2}
  \sum
 \Izerlr_{m_1}(\bar\gamma_{\so}|\bar \alpha)\Izerrl_{m_2}(\bar \beta|\bar\gamma_{\st})f(\bar\gamma_{\st},\bar\gamma_{\so})
 = (-q)^{\pm m_2}f(\bar \beta,\bar\gamma) \Izerlr_{m_1+m_2}(\bar\gamma|\{\bar \alpha,\bar \beta q^2\}).
 \end{equation}
\end{lemma}

This statement is a simple corollary of Lemma~1 of the work \cite{BelPRS12b}.

\section{Properties of $Z_{a,b}$\label{A-PZab}}

The proofs of the properties listed below are given in \cite{PakRS13c}.

\subsection{Sum formulas}
Let $a\ge b$. Then
 \begin{equation}\label{Al-RHC-IHC-twin1}
  \sum
 \Izerrl_b(\bar t_{\so}|\bar y q^2)\Izerlr_b(\bar t_{\so}|\bar sq^2)
 \Izerlr_{a-b}(\bar\xi|\bar t_{\st})f(\bar t_{\st},\bar t_{\so})
   =(-q)^{\pm b}\frac{\RHClr_{a,b}(\bar t;\{\bar\xi,\bar y\}|\bar s;\bar yq^{-2})}{ f(\bar y,\bar t)f(\bar s,\bar t)},
        \end{equation}
 \begin{equation}
  \sum
 \Izerlr_b(\bar t_{\so}|\bar y q^2)\Izerrl_b(\bar t_{\so}|\bar sq^2)
 \Izerlr_{a-b}(\bar\xi|\bar t_{\st})f(\bar t_{\st},\bar t_{\so})
      =(-q)^{\pm b}\frac{\RHClr_{a,b}(\bar t;\{\bar\xi,\bar s\}|\bar y;\bar sq^{-2})}{ f(\bar y,\bar t)f(\bar s,\bar t)}.\label{Al-RHC-IHC-twin2}
     \end{equation}
Here the sum is taken over partitions $\bar t\Rightarrow\{\bar t_{\so},\bar t_{\st}\}$ with
$\#\bar t_{\so}=b$.

\subsection{Reductions of the highest coefficients}

Let $\#\bar z=n$. Then
\be{2-red-1}
\lim_{\bar z'\to\bar z}f^{-1}(\bar z',\bar z)\RHClr_{a+n,b}(\{\bar t,\bar z\};\{\bar x,\bar z'\}|\bar s;\bar y)
=\bar z^{-1}\;f(\bar z,\bar t)f(\bar x,\bar z)f(\bar s,\bar z)\RHClr_{a,b}(\bar t;\bar x|\bar s;\bar y).
\ee
and
\be{2-red-2}
\lim_{\bar z'\to\bar z}f^{-1}(\bar z',\bar z)\RHClr_{a,b+n}(\bar t;\bar x|\{\bar s,\bar z\};\{\bar y,\bar z'\})
=\bar z^{-1}\;f(\bar z,\bar x)f(\bar z,\bar s)f(\bar y,\bar z)\RHClr_{a,b}(\bar t;\bar x|\bar s;\bar y),
\ee

\subsection{Reverse order of arguments}
\be{Z-invers}
 \RHClr_{b,a}(\bar s;\bar y|\bar tq^{-2};\bar xq^{-2})=f^{-1}(\bar y,\bar x)f^{-1}(\bar s,\bar t)
  \RHClr_{a,b}(\bar t;\bar x|\bar s;\bar y).
\ee

\section{Proof of Proposition~\ref{MainProp1}\label{A-PP}}

We start with the
equation \eqref{SP-1-red}. Let us set there $\bar\eta_{\rm i}=\blac$ and $\bar\xi_{\rm i}=\bmub$. This implies
$\bar\eta_0=\blab$ and $\bar\xi_{\rm ii}=\bmuc_{\rm ii}$. We know
that the rational coefficient of the product $r_1(\blac)r_3(\bmub)$ is the highest coefficient $\RHCl$ (up to a normalization):
 \begin{equation}\label{HCl-2}
 S_{a,b}=r_1(\blac)r_3(\bmub)\;
\frac{\RHCl_{a,b}(\blac;\blab|\bmuc;\bmub)}
{f(\bmuc,\blac)f(\bmub,\blab)}+\dots,
   \end{equation}
where dots mean the terms corresponding to all other subsets $\bar\eta_{\rm i}$ and $\bar\xi_{\rm i}$. On the other hand,
we still have the sum over
partitions $\bmuc\Rightarrow\{\bmuc_{\rm i},\bmuc_{\rm ii}\}$ and $\bar\eta_0=\blab\Rightarrow
\{\bar\eta_{\rm ii},\bar\eta_{\rm iii}\}$ in the equation \eqref{SP-1-red}. Thus, we obtain
 \begin{multline}\label{HCl-3}
r_1(\blac)r_3(\bmub)\;
\frac{\RHCl_{a,b}(\blac;\blab|\bmuc;\bmub)}
{f(\bmuc,\blac)f(\bmub,\blab)}=\sum
 \frac{(-q)^{k-b}}{f(\bmuc,\blac)}
 r_1(\blac)r_3(\bmub)\;\Izerr_{b}(\{q^{-2}\bmuc_{\rm ii},q^{-2}\bar\eta_{\rm iii}\}|\bmub)
 \num
\times\RHCl_{a,k}(\blac;\blab|\bmuc_{\rm i};q^{-2}\bar\eta_{\rm iii})\;f(\bar\eta_{\rm ii},\bar\eta_{\rm iii})f(\bmub,\bmuc_{\rm ii})f(\bmuc_{\rm ii},\bmuc_{\rm i})
\num
\times \lim_{\substack{\bar\xi_{\rm ii}\to\bmuc_{\rm ii}\\ \bar\eta_{\rm i}\to\blac}}
\frac{\RHCr_{a,b-k}(\bar\eta_{\rm i};\blac|\bar\xi_{\rm ii};\bmuc_{\rm ii})}
{f(\bmuc_{\rm ii},\blac)f(\bmuc_{\rm ii},\bar\xi_{\rm ii})f(\blac,\bar\eta_{\rm i})} ,
   \end{multline}
where the sum is taken over partitions $\bmuc\Rightarrow\{\bmuc_{\rm i},\bmuc_{\rm ii}\}$ and $\blab\Rightarrow\{\bar\eta_{\rm ii},
\bar\eta_{\rm iii}\}$. Due to \eqref{2-red-2}, \eqref{2-red-1} one has
\be{Evid-2}
\lim_{\substack{\bar\xi_{\rm ii}\to\bmuc_{\rm ii}\\ \bar\eta_{\rm i}\to\blac}}
\frac{\RHCr_{a,b-k}(\bar\eta_{\rm i};\blac|\bar\xi_{\rm ii};\bmuc_{\rm ii})}
{f(\bmuc_{\rm ii},\blac)f(\bmuc_{\rm ii},\bar\xi_{\rm ii})f(\blac,\bar\eta_{\rm i})}=1,
\ee
and we arrive at
 \begin{multline}\label{HCl-4}
f^{-1}(\bmub,\blab)\;\RHCl_{a,b}(\blac;\blab|\bmuc;\bmub)
=\sum
 (-q)^{k-b}\Izerr_{b}(\{q^{-2}\bmuc_{\rm ii},q^{-2}\bar\eta_{\rm iii}\}|\bmub)
 \num
\times\RHCl_{a,k}(\blac;\blab|\bmuc_{\rm i};q^{-2}\bar\eta_{\rm iii})\;
  f(\bar\eta_{\rm ii},\bar\eta_{\rm iii})f(\bmub,\bmuc_{\rm ii})f(\bmuc_{\rm ii},\bmuc_{\rm i}).
   \end{multline}
One can easily recognize in this equation a particular case of the identity \eqref{Sum-F}
corresponding to $\bar z=\emptyset$.

In order to reproduce \eqref{Sum-F} with $\bar z\ne\emptyset$, we replace in \eqref{HCl-4} $a$ by $a+n$
and set there
\be{sets}
\begin{array}{ll}
\blac=\{\bar t,\bar z\},& \bmuc=\bar s,\\
\blab=\{\bar x,\bar z'\},& \bmub=\bar y,
\end{array}
\ee
with $\#\bar t=\#\bar x=a$ and $\#\bar z=\#\bar z'=n$. We obtain
 \begin{multline}\label{Intro-z1}
f^{-1}(\bar y,\bar x)f^{-1}(\bar y,\bar z')\;\RHCl_{a+n,b}(\{\bar t,\bar z\};\{\bar x,\bar z'\}|\bar s;\bar y)
=\sum
 (-q)^{k-b}\Izerr_{b}(\{q^{-2}\bar s_{\st},q^{-2}\bar\eta_{\rm iii}\}|\bar y)
 \num
\times\RHCl_{a+n,k}(\{\bar t,\bar z\};\{\bar x,\bar z'\}|\bar s_{\so};q^{-2}\bar\eta_{\rm iii})\;
  f(\bar\eta_{\rm ii},\bar\eta_{\rm iii})f(\bar y,\bar s_{\st})f(\bar s_{\st},\bar s_{\so}),
   \end{multline}
where the sum is taken over partitions $\bar s\Rightarrow\{\bar s_{\so},\bar s_{\st}\}$ and $\{\bar x,\bar z'\}\Rightarrow\{\bar\eta_{\rm ii},\bar\eta_{\rm iii}\}$.
Dividing both sides of \eqref{Intro-z1} by $f(\bar z',\bar z)$ and taking the limit $\bar z'\to\bar z$
via \eqref{2-red-1} we immediately arrive at
 \begin{multline}\label{Intro-z2}
f^{-1}(\bar y,\bar\eta)\;\RHCl_{a,b}(\bar t;\bar x|\bar s;\bar y)
=\sum
 (-q)^{k-b}\Izerr_{b}(\{q^{-2}\bar s_{\st},q^{-2}\bar\eta_{\rm iii}\}|\bar y)
 \num
\times\RHCl_{a,k}(\bar t;\bar x|\bar s_{\so};q^{-2}\bar\eta_{\rm iii})\;
  f(\bar\eta_{\rm ii},\bar\eta_{\rm iii})f(\bar y,\bar s_{\st})f(\bar s_{\st},\bar s_{\so})f^{-1}(\bar s_{\st},\bar z),
   \end{multline}
where $\bar\eta=\{\bar x,\bar z\}$. Up to notations this is exactly the identity \eqref{Sum-F}.

\end{document}